\documentclass[letterpaper,english,preprint,aps,prapplied,superscriptaddress]{revtex4-2}
\usepackage[T1]{fontenc}
\usepackage[utf8]{inputenc}
\usepackage{color}
\usepackage{babel}
\usepackage{textcomp}
\usepackage{amsmath}
\usepackage{graphicx}
\usepackage[colorlinks=true,bookmarks=false,linkcolor=blue,urlcolor=blue,citecolor=blue]{hyperref}
\usepackage{soul} 
\usepackage{xcolor} 
\usepackage{siunitx}
\newcommand{\SIadj}[2]{\SI[number-unit-product={\text{-}}]{#1}{#2}}

\makeatletter

\pdfpageheight\paperheight
\pdfpagewidth\paperwidth

\makeatother

\begin{document}
	\title{Dynamical backaction in an ultrahigh-finesse fiber-based microcavity}
	\author{Felix Rochau}
	\affiliation{Department of Physics, University of Konstanz, 78457 Konstanz, Germany}
	\author{Irene \surname{S\'{a}nchez Arribas}}
	\affiliation{Department of Physics, University of Konstanz, 78457 Konstanz, Germany}
	\affiliation{present address: Department of Electrical and Computer Engineering, Technical University of Munich, 80333 München, Germany}
	\author{Alexandre Brieussel}
	\affiliation{Department of Physics, University of Konstanz, 78457 Konstanz, Germany}
	\author{Sebastian Stapfner}
	\affiliation{Fakultät für Physik, Ludwig-Maximilians-Universität München, 80539 München, Germany}
	\author{David Hunger}
	\affiliation{Physikalisches Institut, Karlsruher Institut für Technologie, 76131 Karlsruhe, Germany}
	\author{Eva M. Weig}
	\email{eva.weig@tum.de}
	\affiliation{Department of Physics, University of Konstanz, 78457 Konstanz, Germany}
	\affiliation{present address: Department of Electrical and Computer Engineering, Technical University of Munich, 80333 München, Germany}

	\begin{abstract}
		The use of low-dimensional objects in the field of cavity optomechanics is limited by their low scattering cross section compared to the size of the optical cavity mode. Fiber-based Fabry-Pérot microcavities can feature tiny mode cross sections and still maintain a high finesse, boosting the light-matter interaction and thus enabling the sensitive detection of the displacement of minute objects. Here we present such an ultrasensitive microcavity setup with the highest finesse reported so far in loaded fiber cavities, $\mathcal{F} = 195\,000$. We are able to position-tune the static optomechanical coupling to a silicon nitride membrane stripe, reaching frequency pull parameters of up to $\lvert G/2\pi\rvert=\SI{1}{\giga\hertz\per\nano\meter}$. We also demonstrate radiation pressure backaction in the regime of an ultrahigh finesse up to $\mathcal{F}=165\,000$. 
	\end{abstract}
	
	\maketitle

	\section{Introduction}
	
	The prospering field of cavity optomechanics \cite{Aspelmeyer2014} studies the coupling between the vibrations of a macroscopic mechanical object and the mode of an electromagnetic cavity. Possible applications range from exploring quantum signatures of macroscopic objects \cite{Chan2011,Teufel2011}, over generating quantum states of light and matter \cite{Palomaki2013,Lecocq2015,Riedinger2016}, to realizing nonreciprocal devices \cite{Peterson2017,Bernier2017} or acceleration or force sensors \cite{Hutchison2012,Miao2012,Simonsen2019}. Those applications are enabled by the extreme detection sensitivities of optomechanical systems \cite{Mason2019} and often enhanced by minimal thermally induced decoherence at cryogenic temperatures or via  reservoir engineering \cite{Purdy2017,Tebbenjohanns2019}.
	
	A convenient way to  improve the performance of the optomechanical system is to separate the mechanical and the optical cavity mode in a membrane-in-the-middle type configuration \cite{Thompson2008}, which allows to simultaneously optimize both the mechanical resonator and the optical cavity. Typically, the mechanical resonators under investigation are macroscopic objects such as membranes with almost millimeter-scale lateral dimensions \cite{Jayich2008,Wilson2009,Karuza2013,Purdy2013,Reinhardt2016}. However, studying mechanical objects with dimensions in the lower-nanometer regime can be an interesting alternative. As a result of their outstandingly low masses these systems  are extremely sensitive to environmental changes. At the same time they exhibit large zero-point fluctuations $x_\mathrm{zpf}$, enabling large single-photon coupling strengths $g_0$. Optical cavities with small mode volumes \cite{Hunger2010} are perfect candidates to detect the vibrations of such mesoscopic objects. Reducing the mode volume increases the light-matter coupling \cite{Waldron1960}. This boosts the frequency pull parameter  $G=-\frac{\partial\omega_\mathrm{cav}}{\partial z}$, which translates the displacement of the mechanical resonator to a frequency shift of the cavity. Although the frequency pull parameter is set by the geometry and the optical properties of the system, the cavity finesse $\mathcal{F}$ boosts the circulating photon number $n_\mathrm{circ}$ and therefore the effective coupling $g=\sqrt{n_\mathrm{circ}}g_0=\sqrt{n_\mathrm{circ}}Gx_\mathrm{zpf}$. A large finesse also contributes to maintain small cavity linewidths $\kappa = \dfrac{\omega_\mathrm{fsr}}{\mathcal{F}}$ despite the large free spectral range $\omega_\mathrm{fsr}$ of small mode volume cavities. This is necessary to enhance the single photon cooperativity $\mathcal{C}_0 = \dfrac{4g_\mathrm{0}^2}{\kappa\Gamma_\mathrm{m}}$, a key parameter for optomechanical experiments \cite{Aspelmeyer2014}. In addition, the finesse increases the magnitude of phase fluctuations in the output field and, thus, improves the sensitivity of the measurement.

	 The required cavity specifications can be fulfilled with fiber-based Fabry-Pérot microcavities (FFPCs), which have been pioneered in the cavity quantum electrodynamics community \cite{Hunger2010,Muller2010,Gallego2016} but have successfully been adapted to optomechanical \cite{Flowers-Jacobs2012} and ion-trapping systems \cite{Brandstatter2013}. Although FFPC-based optomechanical systems are nowadays being studied by several groups \cite{Stapfner2013,Kashkanova2017,Fogliano2021}, their operation is to date enabled by measures to suppress their extreme sensitivity towards frequency fluctuations. This is conveniently accomplished by  constraining the finesse to not exceed values of a few $10\,000$ and/or by operating at cryogenic temperatures where the Brownian thermal noise of the FFPC mirrors is suppressed.  
	
	In this paper, we introduce a platform for cavity optomechanics that is optimized for ultrasensitive optical detection of single-digit nanometer-size mechanical objects. This is achieved with a FFPC that features a small mode cross section while maintaining the highest finesse reported so far in loaded FFPCs, exceeding that of the ground-breaking work in Ref.~\cite{Flowers-Jacobs2012} by almost one order of magnitude. The stable operation of the cavity is enabled by a very rigid cavity gluing scheme combined with a thorough acoustic shielding. This unlocks the previously unaccessible realm of precision sensing at room temperature using ultrahigh-finesse FFPC systems.

	In the following, we present a proof-of-principle demonstration of the FFPC platform, demonstrating radiation pressure backaction using a free-standing stoichiometric silicon nitride (Si\textsubscript{3}N\textsubscript{4}) membrane stripe which is inserted into the cavity. We note, however, that the choice of the mechanical resonator in this resonator-in-the-middle scheme is flexible, and conveniently allows for the investigation of, for example, tethered membranes \cite{Reinhardt2016,Norte2016}, nanowires \cite{Fogliano2021} or low-dimensional materials such as carbon nanotubes (CNTs) \cite{Favero2008,Stapfner2013,Tavernarakis2018,Barnard2019} or two-dimensional crystals \cite{Singh2014}. In particular, CNTs are discussed as one possible path towards quantum optomechanics at room temperature \cite{Tavernarakis2018}.

	\section{Methods}
	
	\begin{figure}
		\includegraphics{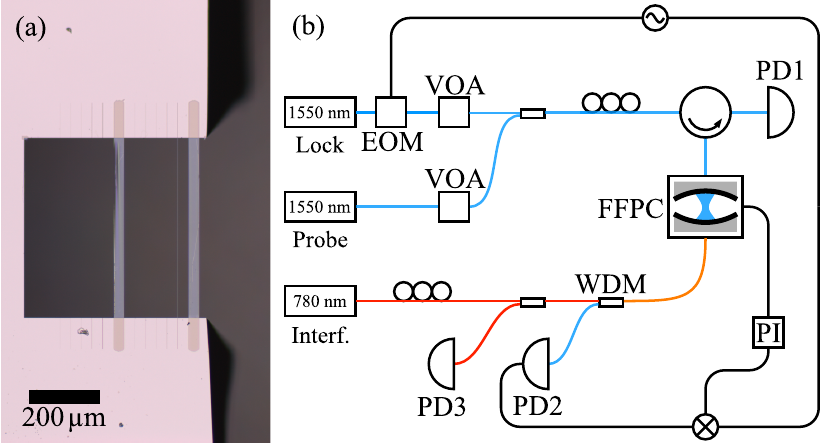} \caption{\label{fig1} Experimental details. (a) Micrograph of the cleaved Si frame (light gray) with two free-standing Si\textsubscript{3}N\textsubscript{4} stripes which have been etched from a $\SI{500}{\micro\meter}$ x $\SI{500}{\micro\meter}$, \SIadj{30}{\nano\meter}-thick stoichiometric membrane. The window opening is cleaved to form a U shape. Parts where the membrane or its supporting frame have been removed appear as a dark gray background. The stripe dimensions are $\SI{30}{\micro\meter}$ x $\SI{500}{\micro\meter}$ x $\SI{30}{\nano\meter}$. (b) Simplified schematic representation of the setup. The cavity length is stabilized on the lock tone's transmission. To this end the lock tone is phase-modulated (EOM). The demodulated transmission from photodetector 2 (PD2) is used as a feedback signal. The probe tone is used to exert a dynamical optical force on the mechanical resonator. The cavity reflection is measured at PD1 using an optical circulator. Both the lock and the probe tone are adjusted in intensity by variable optical attenuators (VOAs). A \SIadj{780}{\nano\meter} interferometric tone is launched from the backside of the cavity (FFPC) and its reflection (PD3) is used to measure the relative sample position.}
	\end{figure}
	
	The measurements presented in this work are performed on a stoichiometric high-stress Si\textsubscript{3}N\textsubscript{4} membrane (Norcada) etched in wide stripes of dimensions $\SI{30}{\micro\meter}\times\SI{500}{\micro\meter}\times\SI{30}{\nano\meter}$. The sample frame is mechanically cleaved to obtain a U shape which allows the membrane stripes to be inserted into the fiber cavity. Figure~\ref{fig1}(a) shows a micrograph of the sample. The mechanical mode of interest is the second harmonic flexural eigenmode ($\mathrm{n}=2$) with a mechanical resonance frequency, linewidth and mechanical quality factor  in vacuum (\SI{e-6}{\milli\bar}) of  $\Omega_0/2\pi= \SI{932.58}{\kilo\hertz}$, $\Gamma_0/2\pi=\SI{4.7}{\hertz}$, and  $Q=197\,000$, respectively. We calculate the effective mass of the sample $\mathrm{m_{eff}} = \SI{0.54}{\nano\gram}$  from its geometry, yielding zero point fluctuations of $x_\mathrm{zpf} = \SI{5.8}{\femto\meter}$. The sample is mounted on a five-axis nanopositioner tower (Attocube ANP101 series) placed close to the FFPC inside a vacuum chamber that enables optical access. To eliminate any source of vibrations the vacuum is maintained at \SI{e-6}{\milli\bar} with an ion pump and the whole chamber is placed inside an acoustically shielded box.

	Figure~\ref{fig1}(b) shows a schematic of the optical setup. A low-noise laser (NKT Koheras Basik X15) with a wavelength of $\SI{1550}{\nano\meter}$ is used as the lock laser for the stabilization of the cavity length. The lock tone is phase-modulated with an electro-optic modulator (EOM) at a frequency of $\omega_{\mathrm{PDH}}/2\pi = \SI{18.26}{\mega\hertz}$. The cavity transmission (PD2) is demodulated in a Poud-Drever-Hall \cite{Drever1983} inspired scheme using a fast lock-in amplifier (Zurich Instruments HF2LI). The error signal is fed to an analog PI controller. The resulting control signal passes a home-built high-voltage amplifier and is sent to shear piezos supporting the fiber mirrors  to stabilize the cavity length. We choose the detuning of the lock tone $\omega_\mathrm{l}$ from the cavity resonance $\omega_\mathrm{cav}$,  $\Delta = \omega_\mathrm{l} - \omega_\mathrm{cav} = 2 \pi \times \SI{150}{\mega\hertz}$  in a way to minimize backaction on the mechanical element. We use a probe tone (NKT Koheras Basik E15) near $\SI{1550}{\nano\meter}$ to measure dynamical backaction by sweeping the probe wavelength across the cavity resonance. A \SIadj{780}{\nano\meter} home-build external cavity diode laser \cite{Ricci1995}  stabilized on a rubidium cell  is used for interferometric readout of the sample position.

	\section{Fiber-Based microcavity}
	
	\begin{figure}
		\includegraphics{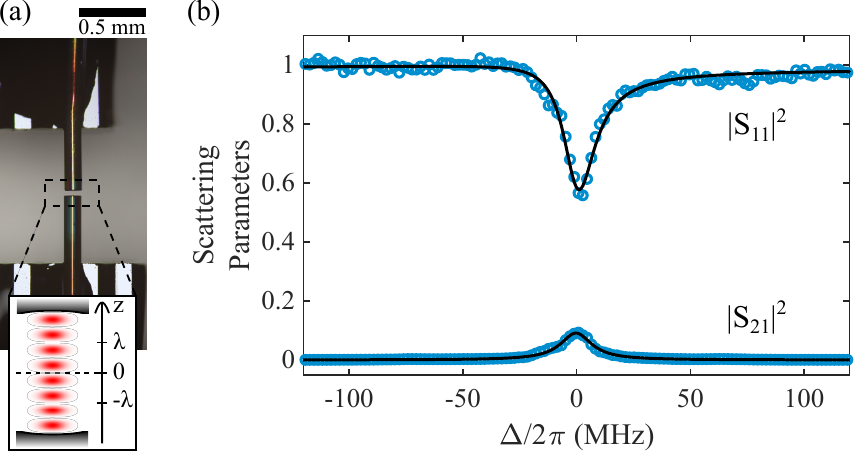} \caption{\label{fig2} Empty fiber cavity. (a) Photograph of the cavity. The fibers are glued to v-groove chips (top and bottom). Inset: schematic of the cavity mode profile. The sample position $z$ is given with respect to its nearest field node. (b) Scattering parameters. Normalized transmission $\lvert S_{21}\rvert ^2$, reflection $\lvert S_{11}\lvert^2$ (blue circles) and fits (black lines) used to extract the loss rates. }
	\end{figure}
	
	The microcavity consists of two mirrors concavely shaped on fiber end faces by CO$_2$ laser ablation, with resulting curvatures of $\SI{191}{\micro\meter}$ and $\SI{140}{\micro\meter}$ for the single-mode (SM) input and multi-mode (MM) output fiber, respectively, which yield a near-planar cavity configuration \cite{Siegman1986}. The mirror spacing is $L = \SI{43.8}{\micro\meter}$. The processed fibers are then ion-beam sputter coated by Laseroptik GmbH with a highly reflective distributed Bragg reflector centered around $\SI{1550}{\nano\meter}$ with a designed transmission of 10\,\si{ppm}. To form the cavity, the fibers are rigidly glued to silicon v-groove chips on top of shear piezo elements. During the gluing process all degrees of freedom are aligned and fixed, resulting in a very stable cavity. By applying a voltage to the shear piezos the cavity length can be tuned by roughly half a wavelength. Figure~\ref{fig2}(a) shows a photograph of the final cavity with the input SM fiber on the top of the picture. 
	
	We use the \SIadj{780}{\nano\meter} laser to measure the relative sample position. This wavelength is outside the coating band of the cavity, allowing a low-finesse interferometric readout of the membrane position. The interferometric tone is launched from the back side of the cavity through the MM fiber. The two wavelengths are split in a wave-division multiplexer (WDM) and the \SIadj{780}{\nano\meter} reflection is measured at PD3 (see Fig. \ref{fig1}(b)).
	
	We characterize the reflection and transmission response of the cavity by fixing the lock laser frequency and scanning the cavity length  across a TEM\textsubscript{00} resonance. Phase-modulated sidebands enable the voltage sent to the cavity piezos to be converted into frequency units. Careful calibration of the gains and losses in the system allows the normalized scattering parameters $\lvert S_{21}\rvert^2=P_\mathrm{t}/P_\mathrm{in}$ and $\lvert S_{11}\rvert ^2=P_\mathrm{r}/P_\mathrm{in}$ to be extracted from the input power $P_\mathrm{in}$ and the power in reflection $P_\mathrm{r}$ and transmission $P_\mathrm{t}$ in front of the cavity (Fig. \ref{fig2}(b)). We fit the normalized transmission and reflection to extract the total cavity losses $\kappa$ and the input couplings $\kappa_\mathrm{e,SM}$ and $\kappa_\mathrm{e,MM}$ of the two mirrors. Due to the imperfect alignment of the mirror surface with respect to the fiber core, light reflected from the cavity is filtered spatially, resulting in an asymmetrical reflection lineshape. This effect is specific to FFPCs and effectively lowers the off-resonant reflection \cite{Gallego2016}. We take the asymmetry into account by adding an heuristic phase factor $V_{bg}\exp{\left(i\phi\right)}$ to $S_{11}$, but a detailed description of this phenomenon can be found in Ref. \cite{Gallego2016}. The final fit formulas read as follows:
	
	\begin{align}\label{eq:eq1}
		\begin{split}
	\lvert S_{11} \rvert ^2 &= \bigg\lvert 1 - \frac{\kappa_\mathrm{e,SM}}{\frac{\kappa}{2}+i\Delta} + V_{bg}\exp{\left(i\phi\right)}\bigg\rvert^2,\\
	\lvert S_{21} \rvert^2 &= \frac{\kappa_\mathrm{e,SM}\kappa_\mathrm{e,MM}}{\left(\frac{\kappa}{2}\right)^2+\Delta^2}.
		\end{split}
	\end{align}

Note that we integrate losses due to imperfect mode matching (especially significant for fiber-based cavities with off-centered mirrors) into the input coupling to comply with the established notation for open cavities. Figure~\ref{fig2}(b) displays the measurements of $\lvert S_{21}\rvert ^2$ and $\lvert S_{11}\rvert ^2$ along with their fits.  The obtained loss rates are $\kappa/2\pi=\SI{16.8}{\mega\hertz}$, $\kappa_\mathrm{e,SM}/2\pi=\SI{2.01}{\mega\hertz}$, and $\kappa_\mathrm{e,MM}/2\pi=\SI{3.17}{\mega\hertz}$. 

We measure a cavity length (mirror spacing)  of $L=\SI{43.8}{\micro\meter}$ by sweeping the probe tone across multiple free spectral ranges. From the cavity length we obtain a free spectral range of $\omega_\mathrm{fsr}/2\pi=\SI{3.42}{\tera\hertz}$. The mode profile can be calculated from the mirror properties. The resulting mode volume is as small as $V\approx277\,\lambda^3$ with a mode waist of $\SI{5.2}{\micro\meter}$. The finesse of the empty cavity is $\mathcal{F}=\dfrac{\omega_\mathrm{fsr}}{\kappa}=204\,000$, which is almost an oder of magnitude higher than  that of Ref.~\cite{Flowers-Jacobs2012}. 

\section{Optomechanical Coupling and Dynamical Backaction}

\begin{figure}
	\includegraphics{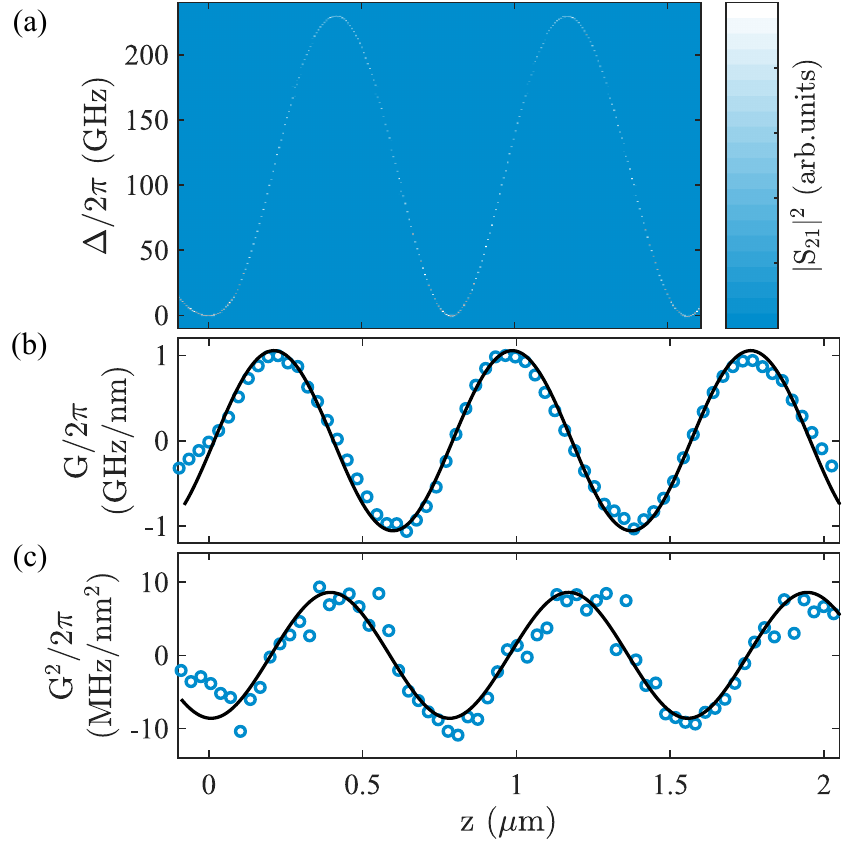} \caption{\label{fig3} Static optomechanical couplings. (a) Normalized cavity transmission $\lvert S_{21}\rvert ^2$ versus  sample position $z$ and detuning. The coordinate $z=0$ corresponds to the sample sitting in an optical field node. (b) Dispersive coupling from measurement (blue circles) and numerical simulation (black line). (c) Quadratic dispersive coupling from measurement (blue circles) and numerical simulation (black line).}
\end{figure}

To map out the dispersive coupling experimentally, that is, the frequency pull parameter $G$, we scan both piezos symmetrically to change the cavity length while the membrane is moved along the cavity axis  \cite{Thompson2008} ($z$ direction, cf. Fig.~\ref{fig2}(a))  and record the cavity transmission and reflection at each position. For this measurement we only use the lock laser at fixed wavelength of \SI{1550}{\nano\meter}. The reflection of the \SIadj{780}{nm} interferometer tone allows us to calibrate the sample position with respect to a certain node of the cavity field. To align the membrane to the cavity mode axis, we maximize the TEM\textsubscript{00} transmission and we minimize scattering into higher-order modes. The map of the cavity detuning $\Delta$ for different sample positions $z$ in Fig. \ref{fig3}(a) reveals a clear sinusoidal periodicity in the intracavity field as expected for a Fabry-Pérot cavity when the mirrors are displaced symmetrically \cite{Thompson2008,Jayich2008}. 

The linear dispersive coupling \cite{Aspelmeyer2014} $G=-\frac{\partial\omega_\mathrm{cav}}{\partial z}$ and the quadratic coupling $G^2=\frac{\partial^2\omega_\mathrm{cav}}{\partial z^2}$ are extracted from this measurement. The resulting linear coupling (Fig. \ref{fig3}(b)) can be as large as $\left|G/2\pi\right|=\SI{1}{\giga\hertz\per\nano\meter}$,  corresponding to a single-photon coupling rate of  $|g_0/2\pi|=\left|G/2\pi\right|\,x_{\mathrm{zpf}}=\SI{5.8}{\kilo\hertz}$. The quadratic coupling is substantial only when the sample is close to the nodes or antinodes of the intracavity field. This is when the quadratic coupling reaches values of up to  $\left|G^2/2\pi\right|=\SI{10}{\mega\hertz\per\nano\meter\squared}$  (Fig. \ref{fig3}(c)).  We can numerically reproduce this modulation via the transfer matrix method \cite{Katsidis2002}. The membrane stripe is simulated as a dielectric slab with a refractive index of $n=1.725+(3.55 \times 10^{-5})i$ (see Appendix \ref{ch:app2}), in accordance with the values reported in the literature \cite{Kim2006, Karuza2012}. The shape and magnitude of both the linear and quadratic coupling agree well with the numerical simulations.

Any absorption in the Si\textsubscript{3}N\textsubscript{4} mechanical resonator at $\SI{1550}{\nano\meter}$, intrinsic and/or due to residues from fabrication,  leads  to a modulation of the cavity losses with respect to the sample position \cite{Jayich2008,Favero2008,Biancofiore2011}. However, this modulation is small due the small value of the imaginary part of the refractive index \cite{Zwickl2008,Karuza2012}. With the sample placed at the node of the cavity, we measure linewidths of \SI{17.5}{\mega\hertz}, corresponding to a loaded finesse of $\mathcal{F}=195\,000$, almost unchanged from the empty cavity. Away from the node, the linewidth stays below \SI{100}{\mega\hertz}, corresponding to $\mathcal{F}>30\,000$. By measuring the linewidth as a function of sample position we can extract the dissipative coupling $G_\kappa = \dfrac{\partial\kappa}{\partial z}$ (see Appendix \ref{ch:app2}) and we obtain a value of the order of  $\rvert G_\kappa/2\pi\lvert = \SI{0.1}{\mega\hertz\per\nano\meter}$. The dissipative coupling is four orders of magnitude smaller than the dispersive coupling. 

We demonstrate dynamical backaction in our system by measuring the optical spring effect \cite{Marquardt2007,WilsonRae2007} from the cavity and the second harmonic of the Si\textsubscript{3}N\textsubscript{4} stripe ($\Omega_0/2\pi = \SI{932.5}{\kilo\hertz}$). For this measurement the sample is placed  $\SI{20}{\nano\meter}$ away from a cavity node, corresponding to $G=\SI{0.18}{\giga\hertz\per\nano\meter}$ or $g_\mathrm{0}=\SI{1}{\kilo\hertz}$ according to the data in Fig.~\ref{fig3}. The cavity length is stabilized on the lock tone and the probe tone is scanned across the cavity resonance. For each probe detuning $\Delta_p$ we record the mechanical power spectral density (PSD), yielding the thermally induced vibrations of the membrane stripe. As their Lorentzian shape is obscured by small yet unavoidable frequency fluctuations, a Voigt fit is employed to extract the mechanical resonance frequency $\mathrm{\Omega_m}$ and effective linewidth $\mathrm{\Gamma_m}$ (see Appendix \ref{ch:app1} for details). 

A detuned drive of the cavity optomechanical system exerts dynamical backaction on the mechanical resonator leading to a modification of the mechanical resonance frequency expressed as $\mathrm{\Omega_m}^2=\Omega_0^2+\delta(\Omega^2)$ (Ref.~\cite{Aspelmeyer2014a}). Including a dissipative coupling term, this frequency shift is given by 

\begin{align}\label{eq:eq3}
\delta(\Omega^2)&=2\Omega_0g^2\left[\frac{\Omega_0+\Delta_p}{(\Omega_0+\Delta_p)^2+\frac{\kappa^2}{4}}-\frac{\Omega_0-\Delta_p}{(\Omega_0-\Delta_p)^2+\frac{\kappa^2}{4}}\right]\nonumber\\
&+2\Omega_0 g g_\kappa \frac{\kappa}{2}\left[\frac{1}{(\Omega_0+\Delta_p)^2+\frac{\kappa^2}{4}}+\frac{1}{(\Omega_0-\Delta_p)^2+\frac{\kappa^2}{4}}\right],
\end{align}

with the effective dispersive coupling $g$ and the effective dissipative coupling $g_\kappa=\sqrt{n_\mathrm{circ}}g_{0,\kappa}$, respectively \cite{Biancofiore2011}.

\begin{figure}
\includegraphics{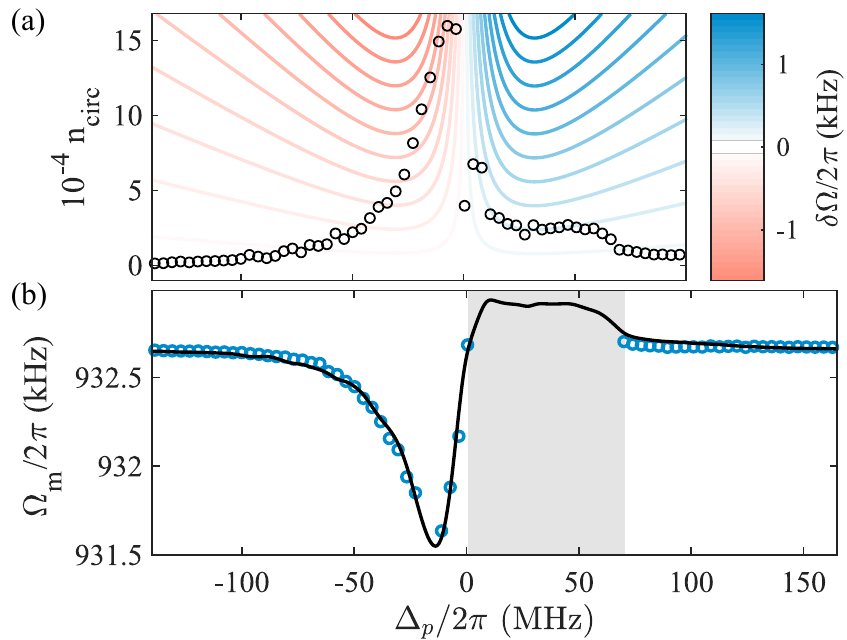}
\caption{Optical spring effect. (a) Calculated mechanical frequency shift (contour lines) and measured circulating photon number (black circles) as a function of the probe detuning with respect to the cavity resonance for a probe input power of $\mathrm{P_{in}} = \SI{7}{\micro\watt}$. (b) Measured mechanical frequency shift (blue circles) of the second harmonic flexural mode and regression (black line). The missing points and gray region correspond to a regime where the resonator undergoes self-sustained oscillations which are unresolved due to the limited measurement bandwidth.}\label{fig4}
\end{figure}

The mechanical frequency shift depends on the circulating photon number through $g = \sqrt{n_{circ}}g_0$ and $g_{\kappa}=\sqrt{n_{circ}}g_{0,\kappa}$, the probe detuning $\Delta_p$, and $\kappa$.  However, the circulating photon number also depends on the probe detuning. In order to be able to fit Equation~\ref{eq:eq3} to the measured mechanical frequency shift, we need to know the exact photon number. To this end, for every detuning step, we calculate the photon number $n_\mathrm{circ}$ from the probe transmission and the launched probe power $P_{in} = \SI{7}{\micro\watt}$ (Fig. \ref{fig4}(a), black circles). For all presented measurements, we sweep the detuning from positive to negative values.   The circulating photon number response appears bistable. First, it increases slightly and at around $\Delta_p = \SI{70}{\mega\hertz}$ the photon number forms a plateau with constant low value. As the detuning approaches zero, the system transitions to a solution with high photon number and the photon number decreases gradually for negative detunings. Here, the response appears Lorentzian as expected, with linewidth $\kappa/2\pi<\SI{21}{\mega\hertz}$, corresponding to $\mathcal{F}>165\,000$.

Figure~\ref{fig4}(b) shows the measured mechanical frequency as a function of the probe detuning. The missing data points for positive detunings correspond to the regime of self-sustained oscillations of the mechanical oscillator, which we are unable to resolve due to the limited bandwidth of the measurement. The black solid line is a fit to Equation \ref{eq:eq3} with best fit parameters $\kappa=\SI{62}{\mega\hertz}$ and $g_0=\SI{575}{\hertz}$, where the measured circulating photon number from Fig.~\ref{fig4} was used as input, and the dissipative coupling was considered negligible. The fit nicely describes the softening of the mode for negative detunings. Around the plateau of the photon number, the mechanical mode oscillates close to its natural frequency. The small stiffening of the mode below a detuning of $\Delta_p = \SI{70}{\mega\hertz}$ is reproduced by the theory. The asymmetry of the fit between negative and positive detunings stems from the large difference in photon numbers. The single-photon coupling strength  extracted from the fit  $\lvert g_0/2\pi\rvert=\SI{575}{\hertz}\,$ is in good agreement with the value of  $\lvert g_0/2\pi\rvert=\SI{1}{\kilo\hertz}$ obtained from the data presented in Fig.~\ref{fig3} at a sample position of \SI{20}{\nano\meter} for both simulations and measurement. The contour lines in Fig.~\ref{fig4}(a)  depict the dependence of the mechanical frequency shift on the circulating photon number and probe detuning for the fit values of  $\kappa=\SI{62}{\mega\hertz}$ and $g_0=\SI{575}{\hertz}$ as from Equation \ref{eq:eq3}.

The cavity linewidth $\kappa = \SI{62}{\mega\hertz}$ that we obtain from the fit in Fig.~\ref{fig4}(b) is three times larger than the cavity	linewidth calculated from the resonant transmission $\kappa = 21\,$MHz. We assume that this is due to noise in the detuning	from cavity length fluctuations. Fluctuations in the cavity length lead to fluctuations of the effective detuning, resulting  in a mismatch between the theoretical and experimental values of the mechanical frequency shift. However, if one neglects photothermal effects inside the mirror coatings and radiation pressure effects on the cavity mirrors, these fluctuations are independent of the photon number. In our measurements we effectively average the mechanical response and the mechanical frequency matches on average the theoretical value (Fig.~\ref{fig4}(b)). This occurs because the fluctuations are symmetric  with respect to the applied effective detuning.

The optical damping arising from dynamical backaction is given by \cite{Biancofiore2011}

\begin{align}\label{eq:eq4}
\Gamma_\mathrm{opt}&=2g^2\frac{\kappa}{2}\left[\frac{1}{(\Omega_0+\Delta_p)^2+\frac{\kappa^2}{4}}-\frac{1}{(\Omega_0-\Delta_p)^2+\frac{\kappa^2}{4}}\right]\nonumber\\
&-2g g_\kappa \left[\frac{\Omega_0-{\Delta_p}}{(\Omega_0-\Delta_p)^2+\frac{\kappa^2}{4}}+\frac{\Omega_0+\Delta_p}{(\Omega_0+\Delta_p)^2+\frac{\kappa^2}{4}}\right],
\end{align}

so that the effective mechanical linewidth $\mathrm{\Gamma_{m}}$  results in $\mathrm{\Gamma_{m} = \Gamma_0 + \Gamma_{opt}}$. In the presence of optomechanical backaction, the thermal force driving the mechanical resonator remains
constant, but the effective mechanical linewidth is altered. To measure the mode temperature, we extract
the effective mechanical damping $\mathrm{\Gamma_{m}}$ from the Voigt fit of the measured mechanical spectra and obtain the effective mode temperature $\mathrm{T_{eff}}$ from the phonon bath temperature $\mathrm{T_{bath}}=\SI{295}{\kelvin}$ via $\mathrm{T_{eff}=T_{bath}\dfrac{\Gamma_0}{\Gamma_{m}}}$.

\begin{figure}
\centering
\includegraphics{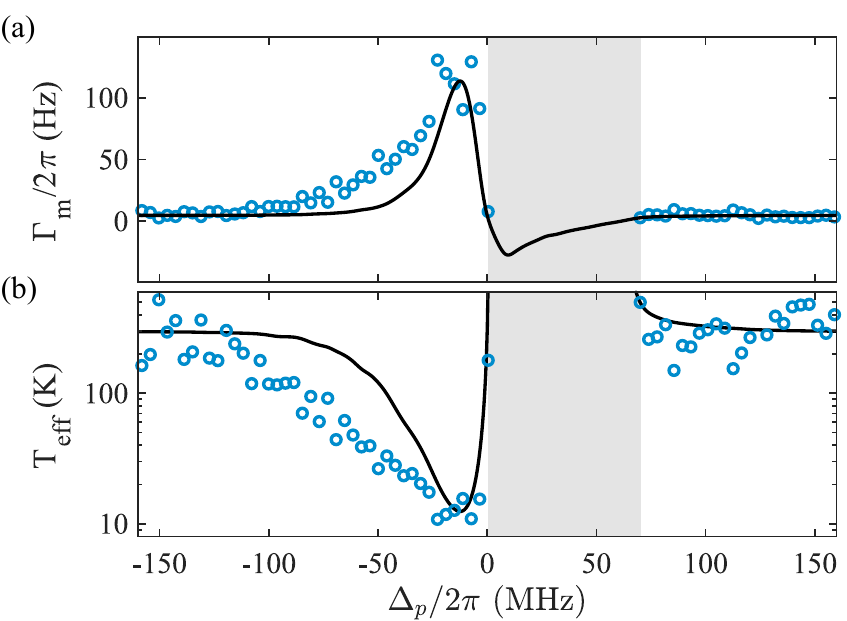}
\caption{Optical damping. (a) Measured effective mechanical
	damping $\mathrm{\Gamma_{m}}$ (blue circles) and theoretical curve (black line) calculated from Equation \ref{eq:eq4} with parameters $g_0 = \SI{575}{\hertz}\,$ and $\kappa = \SI{62}{\mega\hertz}$. The dissipative coupling is neglected in the calculations. (b) Effective mode temperature $\mathrm{T_{eff}=T_{bath}\dfrac{\Gamma_0}{\Gamma_{m}}}$, with $\mathrm{T_{bath}} = \SI{295}{\kelvin}$.  The gray region corresponds to a negative effective mechanical damping.}
\label{fig5}
\end{figure}

 The obtained linewidths are plotted in Fig.~\ref{fig5}(a) together with the theoretical curve (black line) with the parameters from the fit of the optical spring measurement ($g_0 = \SI{575}{\hertz}$, $\kappa = \SI{62}{\mega\hertz}$) and neglecting the dissipative coupling term. Figure~\ref{fig5}(b) displays the effective mode temperature calculated from the measured linewidths. The theory predicts a negative effective mechanical damping, that is, heating of the mechanical mode, for the positive detunings where we measure a plateau of the photon number. In the measurements we observe an effective damping close	to zero which corresponds to self-sustained oscillations. This measurement is limited by the bandwidth of the measurement and the quality of the Voigt profile fits that we use to extract the mechanical linewidth.  For negative detunings, we observe optomechanical cooling of the mode from room temperature down to $\mathrm{T_{eff}}=\SI{12}{K}$, which is verified by additional measurements where we sweep the probe power (see Appendix \ref{ch:app_powersweep}).

\section{Conclusion}

In conclusion, we present an extremely sensitive FFPC setup with a small mode volume and an ultrahigh finesse exceeding $200\,000$. We perform a full characterization of the setup and we demonstrate dynamical backaction on the second harmonic  flexural mode of a stoichiometric Si\textsubscript{3}N\textsubscript{4} membrane stripe. All measurements are supported by theoretical models, allowing  the parameters of the system to be quantitatively extracted. The finesse of the loaded cavity of up to $\mathcal{F}=195\,000$ is exceptionally high for a loaded FFPC system, exceeding that from Ref.~\cite{Flowers-Jacobs2012} by more than one order of magnitude. We report an optomechanical coupling strength of $\lvert g_0/2\pi\rvert=\SI{1}{\kilo\hertz}$ while values up to $\lvert g_0/2\pi\rvert =\SI{5.8}{\kilo\hertz}$ are accessible at a different position of the membrane stripe in the cavity mode. Furthermore, we show self-sustained oscillations of the mechanical mode and optomechanical cooling from room temperature  down to an effective mode temperature of $\mathrm{T_{eff}}=\SI{12}{K}$.

Following the proposal in Ref.~\cite{Favero2008}, the use of single-walled CNTs \cite{Stapfner2013} in this FFPC is very promising. A dispersive coupling strength around $\lvert g_0/2\pi \rvert =\SI{25}{\kilo\hertz}$ is expected, which would place the system deep inside the strong coupling regime.In addition, the outstanding sensitivity of CNTs \cite{Moser2013} enables ultrasensitive cavity optomechanics at room temperature. Owing to the versatility of the presented setup other interesting nanoscale mechanical structures such as nanowires, nanorods, or two-dimensional materials such as h-BN can be studied. By additionally exploiting optical dipole transitions in the said materials, the realization of hybrid optomechanical systems seems to be in reach.

The data supporting the findings of this study are available online \cite{rochau2021}.

\begin{acknowledgments}
Financial support from the Ministry of Research and the Arts of Baden-Württemberg within the Center for Applied Photonics (CAP) is gratefully acknowledged. I.S.A. acknowledges support from the European Union’s Horizon 2020 research and innovation programme under the Marie Sklodowska-Curie grant agreement No. 722923 (OMT).
\end{acknowledgments}

\appendix
\section{\label{ch:app2}Cavity linewidth and dissipative optomechanical coupling}

\begin{figure}
	\centering
	\includegraphics{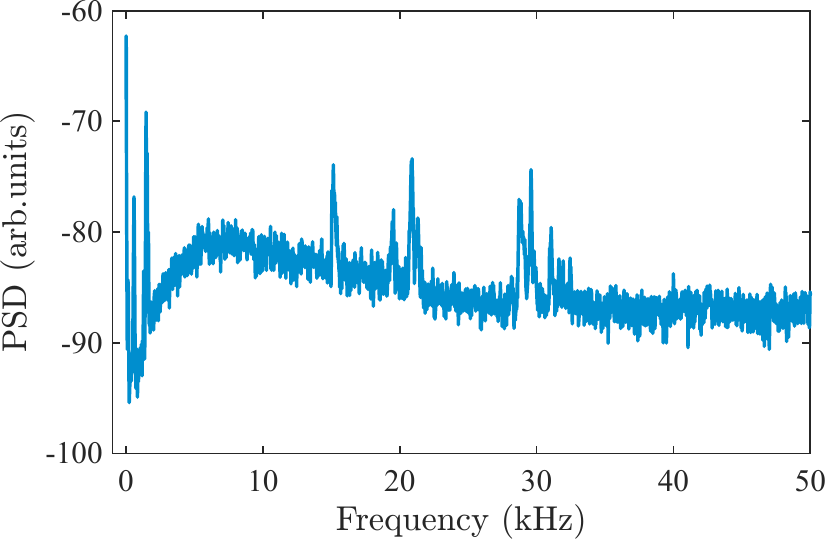}
	\caption{PSD of the locked cavity with sample inserted in arbitrary logarithmic units. The noise is  suppressed due to the cavity stabilization up to around $\SI{5}{\kilo\hertz}$. Residual modes in the feedback band stem from position fluctuations that cannot be compensated by the lock. Peaks above around $\SI{15}{\kilo\hertz}$ are caused by cantilever-type vibrations of the cavity fibers.}
	\label{fig:app2-1}
\end{figure}

The extraction of the cavity linewidth during the static optomechanical coupling measurements is a nontrivial task. Scans over the cavity resonance show a broadened response compared with what should be expected from the normalized on-resonant transmission $\lvert S_{21}\rvert^2$. For positions with a strong dispersive coupling, this broadening originates from the optomechanically induced static bistability \cite{Dorsel1983}. However, even for smaller coupling or small laser powers, fluctuations in the sample position lead to a spectral broadening of the cavity resonance. Those fluctuations arise from acoustical noise and vibrations into the surrounding that couple in the system through the relatively floppy stack of nanopositioners. Vibration damping and acoustic shielding of the setup help to minimize those vibrations in the low-frequency regime (below approximately $\SI{1}{\kilo\hertz}$). The bandwidth of the feedback loop to stabilize the cavity length is limited by resonances of the fiber mirrors acting as cantilever-type mechanical resonators with eigenfrequencies around $\SI{15}{\kilo\hertz}$ (see Fig.~\ref{fig:app2-1}). 

The lifetime of cavity photons of approximately $1/\SI{20}{\mega\hertz}=\SI{50}{\nano\second}$ is much shorter than the timescale of the fluctuations of approximately $1/\SI{20}{\kilo\hertz}=\SI{50}{\micro\second}$, resulting in a spectral response which follows a convolution of a Lorentzian peak  and a Gaussian noise distribution, that is, a Voigt profile (see Appendix~\ref*{ch:app1}). However, the frequency resolution of  the static optomechanical coupling measurements is too small to produce successful Voigt fits due to the large magnitude of the scans. We use instead the linewidth calculated from the normalized on-resonant transmission as a first approximation.  Note that the input power measured on the output port of the circulator and the transmitted power measured on detector PD2 are lower and higher bounds to the actual cavity input and transmission, respectively. Thus, the linewidth that we calculate is an upper bound on the actual cavity linewidth.

\begin{figure}
	\centering
	\includegraphics{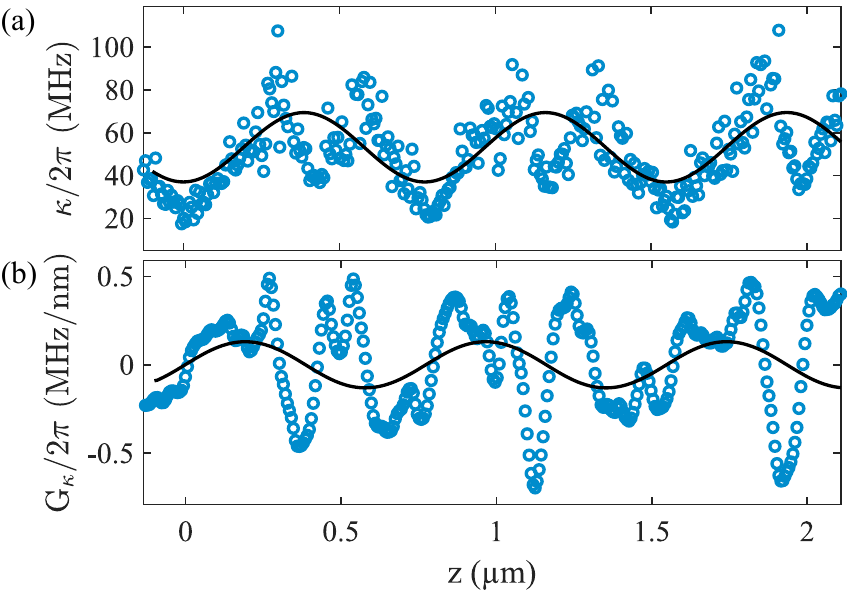}
	\caption{Dissipative optomechanical coupling. (a) Measured (blue circles) and simulated (black line) cavity linewidth with sample inserted. Measured values are upper bounds for the cavity linewidth. (b) Extracted dissipative coupling (blue circles) and numerical simulation (black line).}
	\label{fig:app2}
\end{figure}

Figure~\ref{fig:app2}(a) depicts the measured cavity linewidths with the sample inserted (blue circles) and simulations (black solid line) as a function of the membrane's position.  To reproduce the measurement we need an imaginary part of the refractive index of $\mathrm{Im}(n)=3.55\times 10^{-5}$, which agrees to what has been reported in the literature for stoichiometric Si\textsubscript{3}N\textsubscript{4} (Ref.~\cite{Kim2006, Karuza2012}).  This validates our approximation to extract the cavity linewidth from the normalized on-resonant transmission. 

The dissipative coupling can be calculated from the linewidths obtained both from the measurements and from the numerical simulations (see Fig. \ref{fig:app2}(b)). The exact shape of the measured coupling is hard to extract and depends strongly on how smoothly the derivative is calculated. Nevertheless, both the measurement and simulations show a dissipative coupling of the order of \SI{0.1}{\mega\hertz\per\nano\meter}. Thus, for the measurements presented in this letter, the dissipative coupling is four orders of magnitude smaller than dispersive coupling.

\section{\label{ch:app1}Mechanical Response}

As soon as we lock the cavity, any cavity length fluctuations will alter the detuning and consequently the optical spring effect, which is translated into mechanical frequency noise. As this noise is of Gaussian origin, the resulting mechanical spectra  follows a Voigt distribution - a convolution of a Lorentzian peak with a Gaussian noise distribution - very common in spectroscopy where lines appear Doppler broadened \cite{Olivero1977}. The Voigt profile $V(\omega,\,\sigma,\,\Gamma)$ is given by  
$$V(\omega,\,\sigma,\,\Gamma) = G(\omega,\,\sigma) \star L(\omega,\,\Gamma=2).$$

Here $G(\omega,\,\sigma)$ denotes the Gaussian distribution with standard deviation $\sigma$ and $ L(\omega,\,\Gamma=2)$ is the textbook Cauchy-Lorentz distribution with the half-width at half maximum $\Gamma = 2$. Voigt profiles can also be used to fit peaks that are too narrow to be resolved by the finite bandwidth of the measurement devices \cite{Chen2015}.
\begin{figure}
	\centering
	\includegraphics{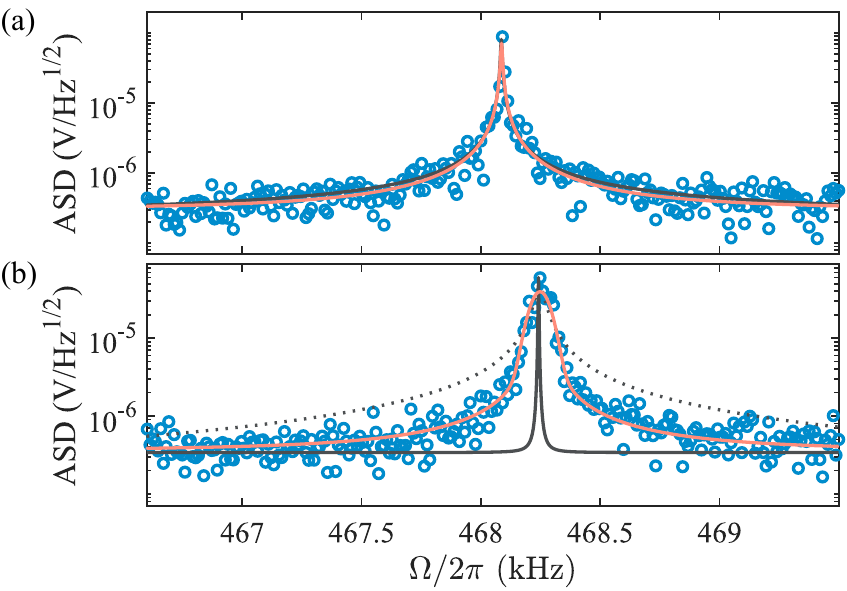}
	\caption{Mechanical peak broadening under cavity backaction noise.
		(a) The spectrum of the Brownian motion of the fundamental harmonic flexural mode can be fitted with both
		a Lorentzian fit (black line) and a Voigt profile fit (orange line). Both regressions yield single-digit
		mechanical dampings of $\mathrm{\Gamma_m}/2\pi = \SI{5.0}{\hertz}$ and $\mathrm{\Gamma_m}/2\pi = \SI{3.3}{\hertz}$, respectively. The Voigt profile yields a Gaussian broadening of  $\sigma/2\pi = \SI{0.1}{\hertz}$. (b) Under weak dynamical backaction, the peak is slightly shifted to higher frequencies and appears broadened. Lorentzian fits (black dotted line fits the central	peak, black solid line fits the tails) cannot describe the resonance shape. A Voigt profile (orange line) fits
		the data and yields $\mathrm{\Gamma_m}/2\pi = \SI{1.7}{\hertz}$ and $\sigma/2\pi = \SI{30}{\hertz}$.}
	\label{fig:app1}
\end{figure}

Figure~\ref{fig:app1} (a) displays the Brownian motion resonance of the fundamental harmonic flexural mode ($n=1$, $\mathrm{\Omega_0/2\pi} = \SI{468.2}{\kilo\hertz}$, $\mathrm{\Gamma_0/2\pi = \SI{10.8}{\hertz}}$,  $\mathrm{Q = 43\,000}$) under dynamical backaction. We can fit the measurement both with a Lorentzian function (black line) and with a
Voigt profile (orange line). Both fits yield low mechanical dampings of $\mathrm{\Gamma_m}/2\pi = \SI{5.0}{\hertz}$ and $\mathrm{\Gamma_m}/2\pi = \SI{3.3}{\hertz}$, respectively. The Voigt profile yields a Gaussian broadening of $\sigma/2\pi = \SI{0.1}{\hertz}$, probably limited by the  resolution bandwidth of the spectrum analyzer. In Fig. \ref{fig:app1} (b) we lock the cavity with the lock tone blue detuned. As expected, the optical spring effect shifts the mechanical resonance to slightly higher frequencies. The expected narrowing of the effective mechanical linewidth is obscured by broadening of
the resonance (Fig. \ref{fig:app1} (b)). A Lorentzian fit of the central part of the peak (black dotted line) largely
overestimates the tails, while a Lorentzian fit of the tails (black solid line) clearly deviates from the
resonance shape. A Voigt profile (orange line) nicely reproduces the data and yields values $\mathrm{\Gamma_m}/2\pi = \SI{1.7}{\hertz}$
and $\sigma/2\pi = \SI{30}{\hertz}$. The extracted effective mechanical damping is reduced from the natural damping as
expected.

\section{Power sweep}\label{ch:app_powersweep}

We perform a power sweep for a fixed probe detuning to confirm our observations of optomechanical cooling. According to Eqs. \ref{eq:eq3} and \ref{eq:eq4} we expect both the mechanical resonance frequency and the effective mechanical linewidth to increase linearly with the number of cavity photons.
In contrast to the measurements in the main text, we consider the fundamental harmonic flexural mode in the following ($n = 1$, $\mathrm{\Omega_0/2\pi} = \SI{468.2}{\kilo\hertz}$, $\mathrm{\Gamma_0/2\pi = \SI{10.8}{\hertz}}$,  $\mathrm{Q = 43\,000}$). As a result of its lower mechanical quality factor this mode is less prone to start self-oscillating. We place the sample
at a position of around $\mathrm{z} = -\SI{30}{\nano\meter}$ (below a node of the cavity field) resulting in a cavity linewidth of $\kappa/2\pi = \SI{22.3}{\mega\hertz}$ and stabilize the cavity in a way that the lock tone is blue detuned.

We fix a probe detuning of $\Delta_p/\kappa = 0.56$  and we scan the probe power from $\mathrm{P_{in} = \SI{0.2}{\micro\watt}}$ to $\mathrm{P_{in} = \SI{18}{\micro\watt}}$. That corresponds to a number of circulating photons due to the probe tone of up to $\mathrm{n_{circ} = 160\,000}$. The photon number due to the lock tone is constant during the measurement, $\mathrm{n_{circ} \sim 90\,000}$. We select a large lock detuning to minimize backaction from the lock tone. At the same time, a large lock photon number ensures a strong feedback signal that boosts the lock quality.

We record the mechanical spectra of both the \SIadj{780}{\nano\meter} interferometer and the X quadrature extracted from the heterodyne signal in transmission. The mechanical PSDs are fitted with Voigt profiles to obtain the mechanical resonances and the effective mechanical dampings.

\begin{figure}
	\centering
	\includegraphics{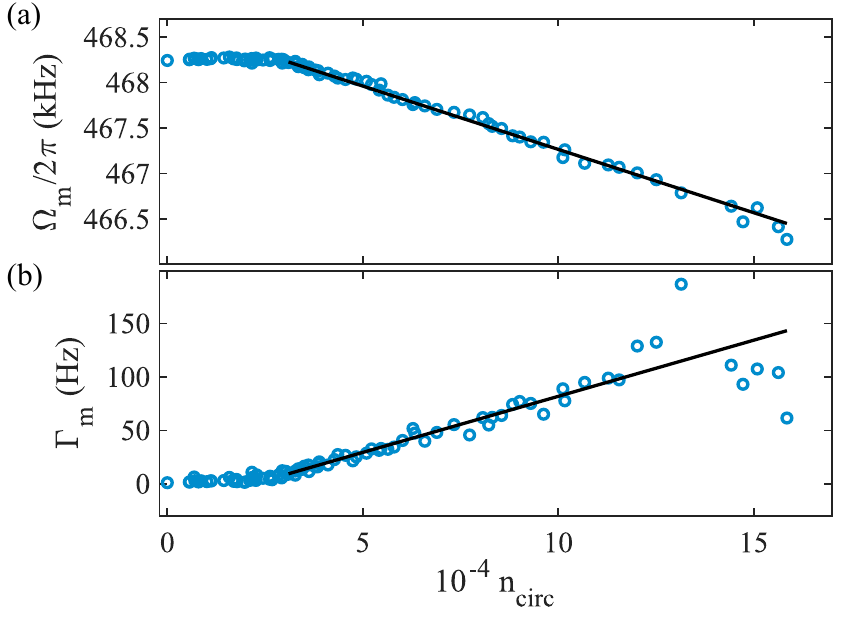}
	\caption{Dynamical backaction cooling. (a) The mechanical resonance frequency $\mathrm{\Omega_m}$ decreases linearly with increasing photon number of the probe tone $\mathrm{n_{circ}}$. A linear regression (black line) yields a single-photon coupling of $g_0/2\pi =\SI{395}{\hertz}$. (b) The effective mechanical damping $\mathrm{\Gamma_m}$ rises linearly with	increasing photon number. For very high probe powers the damping decreases again. A regression of the linear part (black line) yields $g_0/2\pi=\SI{401}{\hertz}$.}	
	\label{fig:app_powersweep}
	
\end{figure}

Figure~\ref{fig:app_powersweep}(a) displays the mechanical frequency as a function of probe photon number. From around $\mathrm{n_{circ} = 30\,000}$ the mechanical resonance decreases linearly with the probe photon number. The slope of a linear regression (black line) in this range yields a dispersive single-photon coupling of $g_0/2\pi=\SI{395(5)}{\hertz}$. If we consider the effective mechanical damping in  Fig.~\ref{fig:app_powersweep}(b) we observe
the expected linear increase of $\mathrm{\Gamma_m}$ with the photon number in the range between  $\mathrm{n_{circ} = 30\,000}$ and
 $\mathrm{n_{circ} = 130\,000}$. Again, we extract the single-photon coupling from a linear regression and we obtain $g_0/2\pi=\SI{401(10)}{\hertz}$. Those values appear reasonable compared with the single-photon couplings $g_0/2\pi=\SI{575}{\hertz}$ from the main text.
 
The mode self-oscillates for  photon numbers below $\mathrm{n_{circ}}=30\,000$, where the mechanical resonance frequency remains constant and the  observed effective damping is close to zero. The linear regression allows to estimate the optical damping  induced by the lock tone to be of the order of $\mathrm{\Gamma_{opt}/2\pi} = \SI{-40}{\hertz}$.

\clearpage


\begin{thebibliography}{50}%
	\makeatletter
	\providecommand \@ifxundefined [1]{%
		\@ifx{#1\undefined}
	}%
	\providecommand \@ifnum [1]{%
		\ifnum #1\expandafter \@firstoftwo
		\else \expandafter \@secondoftwo
		\fi
	}%
	\providecommand \@ifx [1]{%
		\ifx #1\expandafter \@firstoftwo
		\else \expandafter \@secondoftwo
		\fi
	}%
	\providecommand \natexlab [1]{#1}%
	\providecommand \enquote  [1]{``#1''}%
	\providecommand \bibnamefont  [1]{#1}%
	\providecommand \bibfnamefont [1]{#1}%
	\providecommand \citenamefont [1]{#1}%
	\providecommand \href@noop [0]{\@secondoftwo}%
	\providecommand \href [0]{\begingroup \@sanitize@url \@href}%
	\providecommand \@href[1]{\@@startlink{#1}\@@href}%
	\providecommand \@@href[1]{\endgroup#1\@@endlink}%
	\providecommand \@sanitize@url [0]{\catcode `\\12\catcode `\$12\catcode
		`\&12\catcode `\#12\catcode `\^12\catcode `\_12\catcode `\%12\relax}%
	\providecommand \@@startlink[1]{}%
	\providecommand \@@endlink[0]{}%
	\providecommand \url  [0]{\begingroup\@sanitize@url \@url }%
	\providecommand \@url [1]{\endgroup\@href {#1}{\urlprefix }}%
	\providecommand \urlprefix  [0]{URL }%
	\providecommand \Eprint [0]{\href }%
	\providecommand \doibase [0]{https://doi.org/}%
	\providecommand \selectlanguage [0]{\@gobble}%
	\providecommand \bibinfo  [0]{\@secondoftwo}%
	\providecommand \bibfield  [0]{\@secondoftwo}%
	\providecommand \translation [1]{[#1]}%
	\providecommand \BibitemOpen [0]{}%
	\providecommand \bibitemStop [0]{}%
	\providecommand \bibitemNoStop [0]{.\EOS\space}%
	\providecommand \EOS [0]{\spacefactor3000\relax}%
	\providecommand \BibitemShut  [1]{\csname bibitem#1\endcsname}%
	\let\auto@bib@innerbib\@empty
	\bibitem [{\citenamefont {Aspelmeyer}\ \emph
		{et~al.}(2014{\natexlab{a}})\citenamefont {Aspelmeyer}, \citenamefont
		{Kippenberg},\ and\ \citenamefont {Marquardt}}]{Aspelmeyer2014}%
	\BibitemOpen
	\bibfield  {author} {\bibinfo {author} {\bibfnamefont {M.}~\bibnamefont
			{Aspelmeyer}}, \bibinfo {author} {\bibfnamefont {T.~J.}\ \bibnamefont
			{Kippenberg}},\ and\ \bibinfo {author} {\bibfnamefont {F.}~\bibnamefont
			{Marquardt}},\ }\bibfield  {title} {\bibinfo {title} {Cavity optomechanics},\
	}\href {https://doi.org/10.1103/revmodphys.86.1391} {\bibfield  {journal}
		{\bibinfo  {journal} {Rev. Mod. Phys.}\ }\textbf {\bibinfo {volume} {86}},\
		\bibinfo {pages} {1391} (\bibinfo {year} {2014}{\natexlab{a}})}\BibitemShut
	{NoStop}%
	\bibitem [{\citenamefont {Chan}\ \emph {et~al.}(2011)\citenamefont {Chan},
		\citenamefont {Alegre}, \citenamefont {Safavi-Naeini}, \citenamefont {Hill},
		\citenamefont {Krause}, \citenamefont {Gröblacher}, \citenamefont
		{Aspelmeyer},\ and\ \citenamefont {Painter}}]{Chan2011}%
	\BibitemOpen
	\bibfield  {author} {\bibinfo {author} {\bibfnamefont {J.}~\bibnamefont
			{Chan}}, \bibinfo {author} {\bibfnamefont {T.~P.~M.}\ \bibnamefont {Alegre}},
		\bibinfo {author} {\bibfnamefont {A.~H.}\ \bibnamefont {Safavi-Naeini}},
		\bibinfo {author} {\bibfnamefont {J.~T.}\ \bibnamefont {Hill}}, \bibinfo
		{author} {\bibfnamefont {A.}~\bibnamefont {Krause}}, \bibinfo {author}
		{\bibfnamefont {S.}~\bibnamefont {Gröblacher}}, \bibinfo {author}
		{\bibfnamefont {M.}~\bibnamefont {Aspelmeyer}},\ and\ \bibinfo {author}
		{\bibfnamefont {O.}~\bibnamefont {Painter}},\ }\bibfield  {title} {\bibinfo
		{title} {Laser cooling of a nanomechanical oscillator into its quantum ground
			state},\ }\href {https://doi.org/10.1038/nature10461} {\bibfield  {journal}
		{\bibinfo  {journal} {Nature}\ }\textbf {\bibinfo {volume} {478}},\ \bibinfo
		{pages} {89} (\bibinfo {year} {2011})}\BibitemShut {NoStop}%
	\bibitem [{\citenamefont {Teufel}\ \emph {et~al.}(2011)\citenamefont {Teufel},
		\citenamefont {Donner}, \citenamefont {Li}, \citenamefont {Harlow},
		\citenamefont {Allman}, \citenamefont {Cicak}, \citenamefont {Sirois},
		\citenamefont {Whittaker}, \citenamefont {Lehnert},\ and\ \citenamefont
		{Simmonds}}]{Teufel2011}%
	\BibitemOpen
	\bibfield  {author} {\bibinfo {author} {\bibfnamefont {J.~D.}\ \bibnamefont
			{Teufel}}, \bibinfo {author} {\bibfnamefont {T.}~\bibnamefont {Donner}},
		\bibinfo {author} {\bibfnamefont {D.}~\bibnamefont {Li}}, \bibinfo {author}
		{\bibfnamefont {J.~W.}\ \bibnamefont {Harlow}}, \bibinfo {author}
		{\bibfnamefont {M.~S.}\ \bibnamefont {Allman}}, \bibinfo {author}
		{\bibfnamefont {K.}~\bibnamefont {Cicak}}, \bibinfo {author} {\bibfnamefont
			{A.~J.}\ \bibnamefont {Sirois}}, \bibinfo {author} {\bibfnamefont {J.~D.}\
			\bibnamefont {Whittaker}}, \bibinfo {author} {\bibfnamefont {K.~W.}\
			\bibnamefont {Lehnert}},\ and\ \bibinfo {author} {\bibfnamefont {R.~W.}\
			\bibnamefont {Simmonds}},\ }\bibfield  {title} {\bibinfo {title} {Sideband
			cooling of micromechanical motion to the quantum ground state},\ }\href
	{https://doi.org/10.1038/nature10261} {\bibfield  {journal} {\bibinfo
			{journal} {Nature}\ }\textbf {\bibinfo {volume} {475}},\ \bibinfo {pages}
		{359} (\bibinfo {year} {2011})}\BibitemShut {NoStop}%
	\bibitem [{\citenamefont {Palomaki}\ \emph {et~al.}(2013)\citenamefont
		{Palomaki}, \citenamefont {Teufel}, \citenamefont {Simmonds},\ and\
		\citenamefont {Lehnert}}]{Palomaki2013}%
	\BibitemOpen
	\bibfield  {author} {\bibinfo {author} {\bibfnamefont {T.~A.}\ \bibnamefont
			{Palomaki}}, \bibinfo {author} {\bibfnamefont {J.~D.}\ \bibnamefont
			{Teufel}}, \bibinfo {author} {\bibfnamefont {R.~W.}\ \bibnamefont
			{Simmonds}},\ and\ \bibinfo {author} {\bibfnamefont {K.~W.}\ \bibnamefont
			{Lehnert}},\ }\bibfield  {title} {\bibinfo {title} {Entangling mechanical
			motion with microwave fields},\ }\href
	{https://doi.org/10.1126/science.1244563} {\bibfield  {journal} {\bibinfo
			{journal} {Science}\ }\textbf {\bibinfo {volume} {342}},\ \bibinfo {pages}
		{710} (\bibinfo {year} {2013})}\BibitemShut {NoStop}%
	\bibitem [{\citenamefont {Lecocq}\ \emph {et~al.}(2015)\citenamefont {Lecocq},
		\citenamefont {Clark}, \citenamefont {Simmonds}, \citenamefont {Aumentado},\
		and\ \citenamefont {Teufel}}]{Lecocq2015}%
	\BibitemOpen
	\bibfield  {author} {\bibinfo {author} {\bibfnamefont {F.}~\bibnamefont
			{Lecocq}}, \bibinfo {author} {\bibfnamefont {J.~B.}\ \bibnamefont {Clark}},
		\bibinfo {author} {\bibfnamefont {R.~W.}\ \bibnamefont {Simmonds}}, \bibinfo
		{author} {\bibfnamefont {J.}~\bibnamefont {Aumentado}},\ and\ \bibinfo
		{author} {\bibfnamefont {J.~D.}\ \bibnamefont {Teufel}},\ }\bibfield  {title}
	{\bibinfo {title} {Quantum nondemolition measurement of a nonclassical state
			of a massive object},\ }\href {https://doi.org/10.1103/PhysRevX.5.041037}
	{\bibfield  {journal} {\bibinfo  {journal} {Phys. Rev. X}\ }\textbf {\bibinfo
			{volume} {5}},\ \bibinfo {pages} {041037} (\bibinfo {year}
		{2015})}\BibitemShut {NoStop}%
	\bibitem [{\citenamefont {Riedinger}\ \emph {et~al.}(2016)\citenamefont
		{Riedinger}, \citenamefont {Hong}, \citenamefont {A.}, \citenamefont
		{Slater}, \citenamefont {Shang}, \citenamefont {Krause}, \citenamefont
		{Anant}, \citenamefont {Aspelmeyer},\ and\ \citenamefont
		{Gröblacher}}]{Riedinger2016}%
	\BibitemOpen
	\bibfield  {author} {\bibinfo {author} {\bibfnamefont {R.}~\bibnamefont
			{Riedinger}}, \bibinfo {author} {\bibfnamefont {S.}~\bibnamefont {Hong}},
		\bibinfo {author} {\bibfnamefont {R.}~\bibnamefont {A.}}, \bibinfo {author}
		{\bibfnamefont {J.~A.}\ \bibnamefont {Slater}}, \bibinfo {author}
		{\bibfnamefont {J.}~\bibnamefont {Shang}}, \bibinfo {author} {\bibfnamefont
			{A.~G.}\ \bibnamefont {Krause}}, \bibinfo {author} {\bibfnamefont
			{V.}~\bibnamefont {Anant}}, \bibinfo {author} {\bibfnamefont
			{M.}~\bibnamefont {Aspelmeyer}},\ and\ \bibinfo {author} {\bibfnamefont
			{S.}~\bibnamefont {Gröblacher}},\ }\bibfield  {title} {\bibinfo {title}
		{Non-classical correlations between single photons and phonons from a
			mechanical oscillator},\ }\href {https://doi.org/10.1038/nature16536}
	{\bibfield  {journal} {\bibinfo  {journal} {Nature}\ }\textbf {\bibinfo
			{volume} {530}},\ \bibinfo {pages} {313} (\bibinfo {year}
		{2016})}\BibitemShut {NoStop}%
	\bibitem [{\citenamefont {Peterson}\ \emph {et~al.}(2017)\citenamefont
		{Peterson}, \citenamefont {Lecocq}, \citenamefont {Cicak}, \citenamefont
		{Simmonds}, \citenamefont {Aumentado},\ and\ \citenamefont
		{Teufel}}]{Peterson2017}%
	\BibitemOpen
	\bibfield  {author} {\bibinfo {author} {\bibfnamefont {G.~A.}\ \bibnamefont
			{Peterson}}, \bibinfo {author} {\bibfnamefont {F.}~\bibnamefont {Lecocq}},
		\bibinfo {author} {\bibfnamefont {K.}~\bibnamefont {Cicak}}, \bibinfo
		{author} {\bibfnamefont {R.~W.}\ \bibnamefont {Simmonds}}, \bibinfo {author}
		{\bibfnamefont {J.}~\bibnamefont {Aumentado}},\ and\ \bibinfo {author}
		{\bibfnamefont {J.~D.}\ \bibnamefont {Teufel}},\ }\bibfield  {title}
	{\bibinfo {title} {Demonstration of efficient nonreciprocity in a microwave
			optomechanical circuit},\ }\href {https://doi.org/10.1103/PhysRevX.7.031001}
	{\bibfield  {journal} {\bibinfo  {journal} {Phys. Rev. X}\ }\textbf {\bibinfo
			{volume} {7}},\ \bibinfo {pages} {031001} (\bibinfo {year}
		{2017})}\BibitemShut {NoStop}%
	\bibitem [{\citenamefont {Bernier}\ \emph {et~al.}(2017)\citenamefont
		{Bernier}, \citenamefont {T{\'{o}}th}, \citenamefont {Koottandavida},
		\citenamefont {Ioannou}, \citenamefont {Malz}, \citenamefont {Nunnenkamp},
		\citenamefont {Feofanov},\ and\ \citenamefont {Kippenberg}}]{Bernier2017}%
	\BibitemOpen
	\bibfield  {author} {\bibinfo {author} {\bibfnamefont {N.~R.}\ \bibnamefont
			{Bernier}}, \bibinfo {author} {\bibfnamefont {L.~D.}\ \bibnamefont
			{T{\'{o}}th}}, \bibinfo {author} {\bibfnamefont {A.}~\bibnamefont
			{Koottandavida}}, \bibinfo {author} {\bibfnamefont {M.~A.}\ \bibnamefont
			{Ioannou}}, \bibinfo {author} {\bibfnamefont {D.}~\bibnamefont {Malz}},
		\bibinfo {author} {\bibfnamefont {A.}~\bibnamefont {Nunnenkamp}}, \bibinfo
		{author} {\bibfnamefont {A.~K.}\ \bibnamefont {Feofanov}},\ and\ \bibinfo
		{author} {\bibfnamefont {T.~J.}\ \bibnamefont {Kippenberg}},\ }\bibfield
	{title} {\bibinfo {title} {Nonreciprocal reconfigurable microwave
			optomechanical circuit},\ }\href {https://doi.org/10.1038/s41467-017-00447-1}
	{\bibfield  {journal} {\bibinfo  {journal} {Nat. Commun.}\ }\textbf {\bibinfo
			{volume} {8}},\ \bibinfo {pages} {604} (\bibinfo {year} {2017})}\BibitemShut
	{NoStop}%
	\bibitem [{\citenamefont {Hutchison}\ and\ \citenamefont
		{Bhave}(2012)}]{Hutchison2012}%
	\BibitemOpen
	\bibfield  {author} {\bibinfo {author} {\bibfnamefont {D.~N.}\ \bibnamefont
			{Hutchison}}\ and\ \bibinfo {author} {\bibfnamefont {S.~A.}\ \bibnamefont
			{Bhave}},\ }\bibfield  {title} {\bibinfo {title} {Z-axis optomechanical
			accelerometer},\ }in\ \href {https://doi.org/10.1109/memsys.2012.6170263}
	{\emph {\bibinfo {booktitle} {2012 {IEEE} 25th International Conference on
				Micro Electro Mechanical Systems ({MEMS})}}}\ (\bibinfo  {publisher}
	{{IEEE}},\ \bibinfo {year} {2012})\BibitemShut {NoStop}%
	\bibitem [{\citenamefont {Miao}\ \emph {et~al.}(2012)\citenamefont {Miao},
		\citenamefont {Srinivasan},\ and\ \citenamefont {Aksyuk}}]{Miao2012}%
	\BibitemOpen
	\bibfield  {author} {\bibinfo {author} {\bibfnamefont {H.}~\bibnamefont
			{Miao}}, \bibinfo {author} {\bibfnamefont {K.}~\bibnamefont {Srinivasan}},\
		and\ \bibinfo {author} {\bibfnamefont {V.}~\bibnamefont {Aksyuk}},\
	}\bibfield  {title} {\bibinfo {title} {A microelectromechanically controlled
			cavity optomechanical sensing system},\ }\href
	{https://doi.org/10.1088/1367-2630/14/7/075015} {\bibfield  {journal}
		{\bibinfo  {journal} {New J. Phys.}\ }\textbf {\bibinfo {volume} {14}},\
		\bibinfo {pages} {075015} (\bibinfo {year} {2012})}\BibitemShut {NoStop}%
	\bibitem [{\citenamefont {Simonsen}\ \emph {et~al.}(2019)\citenamefont
		{Simonsen}, \citenamefont {S{\'{a}}nchez-Heredia}, \citenamefont {Saarinen},
		\citenamefont {Ardenkj{\ae}r-Larsen}, \citenamefont {Schliesser},\ and\
		\citenamefont {Polzik}}]{Simonsen2019}%
	\BibitemOpen
	\bibfield  {author} {\bibinfo {author} {\bibfnamefont {A.}~\bibnamefont
			{Simonsen}}, \bibinfo {author} {\bibfnamefont {J.~D.}\ \bibnamefont
			{S{\'{a}}nchez-Heredia}}, \bibinfo {author} {\bibfnamefont {S.~A.}\
			\bibnamefont {Saarinen}}, \bibinfo {author} {\bibfnamefont {J.~H.}\
			\bibnamefont {Ardenkj{\ae}r-Larsen}}, \bibinfo {author} {\bibfnamefont
			{A.}~\bibnamefont {Schliesser}},\ and\ \bibinfo {author} {\bibfnamefont
			{E.~S.}\ \bibnamefont {Polzik}},\ }\bibfield  {title} {\bibinfo {title}
		{Magnetic resonance imaging with optical preamplification and detection},\
	}\href {https://doi.org/10.1038/s41598-019-54200-3} {\bibfield  {journal}
		{\bibinfo  {journal} {Sci. Rep.}\ }\textbf {\bibinfo {volume} {9}},\ \bibinfo
		{pages} {18173} (\bibinfo {year} {2019})}\BibitemShut {NoStop}%
	\bibitem [{\citenamefont {Mason}\ \emph {et~al.}(2019)\citenamefont {Mason},
		\citenamefont {Chen}, \citenamefont {Rossi}, \citenamefont {Tsaturyan},\ and\
		\citenamefont {Schliesser}}]{Mason2019}%
	\BibitemOpen
	\bibfield  {author} {\bibinfo {author} {\bibfnamefont {D.}~\bibnamefont
			{Mason}}, \bibinfo {author} {\bibfnamefont {J.}~\bibnamefont {Chen}},
		\bibinfo {author} {\bibfnamefont {M.}~\bibnamefont {Rossi}}, \bibinfo
		{author} {\bibfnamefont {Y.}~\bibnamefont {Tsaturyan}},\ and\ \bibinfo
		{author} {\bibfnamefont {A.}~\bibnamefont {Schliesser}},\ }\bibfield  {title}
	{\bibinfo {title} {Continuous force and displacement measurement below the
			standard quantum limit},\ }\href {https://doi.org/10.1038/s41567-019-0533-5}
	{\bibfield  {journal} {\bibinfo  {journal} {Nat. Phys.}\ }\textbf {\bibinfo
			{volume} {15}},\ \bibinfo {pages} {745} (\bibinfo {year} {2019})}\BibitemShut
	{NoStop}%
	\bibitem [{\citenamefont {Purdy}\ \emph {et~al.}(2017)\citenamefont {Purdy},
		\citenamefont {Grutter}, \citenamefont {Srinivasan},\ and\ \citenamefont
		{Taylor}}]{Purdy2017}%
	\BibitemOpen
	\bibfield  {author} {\bibinfo {author} {\bibfnamefont {T.~P.}\ \bibnamefont
			{Purdy}}, \bibinfo {author} {\bibfnamefont {K.~E.}\ \bibnamefont {Grutter}},
		\bibinfo {author} {\bibfnamefont {K.}~\bibnamefont {Srinivasan}},\ and\
		\bibinfo {author} {\bibfnamefont {J.~M.}\ \bibnamefont {Taylor}},\ }\bibfield
	{title} {\bibinfo {title} {Quantum correlations from a room-temperature
			optomechanical cavity},\ }\href {https://doi.org/10.1126/science.aag1407}
	{\bibfield  {journal} {\bibinfo  {journal} {Science}\ }\textbf {\bibinfo
			{volume} {356}},\ \bibinfo {pages} {1265} (\bibinfo {year}
		{2017})}\BibitemShut {NoStop}%
	\bibitem [{\citenamefont {Tebbenjohanns}\ \emph {et~al.}(2019)\citenamefont
		{Tebbenjohanns}, \citenamefont {Frimmer}, \citenamefont {Militaru},
		\citenamefont {Jain},\ and\ \citenamefont {Novotny}}]{Tebbenjohanns2019}%
	\BibitemOpen
	\bibfield  {author} {\bibinfo {author} {\bibfnamefont {F.}~\bibnamefont
			{Tebbenjohanns}}, \bibinfo {author} {\bibfnamefont {M.}~\bibnamefont
			{Frimmer}}, \bibinfo {author} {\bibfnamefont {A.}~\bibnamefont {Militaru}},
		\bibinfo {author} {\bibfnamefont {V.}~\bibnamefont {Jain}},\ and\ \bibinfo
		{author} {\bibfnamefont {L.}~\bibnamefont {Novotny}},\ }\bibfield  {title}
	{\bibinfo {title} {Cold damping of an optically levitated nanoparticle to
			microkelvin temperatures},\ }\href
	{https://doi.org/10.1103/PhysRevLett.122.223601} {\bibfield  {journal}
		{\bibinfo  {journal} {Phys. Rev. Lett.}\ }\textbf {\bibinfo {volume} {122}},\
		\bibinfo {pages} {223601} (\bibinfo {year} {2019})}\BibitemShut {NoStop}%
	\bibitem [{\citenamefont {Thompson}\ \emph {et~al.}(2008)\citenamefont
		{Thompson}, \citenamefont {Zwickl}, \citenamefont {Jayich}, \citenamefont
		{Marquardt}, \citenamefont {Girvin},\ and\ \citenamefont
		{Harris}}]{Thompson2008}%
	\BibitemOpen
	\bibfield  {author} {\bibinfo {author} {\bibfnamefont {J.~D.}\ \bibnamefont
			{Thompson}}, \bibinfo {author} {\bibfnamefont {B.~M.}\ \bibnamefont
			{Zwickl}}, \bibinfo {author} {\bibfnamefont {A.~M.}\ \bibnamefont {Jayich}},
		\bibinfo {author} {\bibfnamefont {F.}~\bibnamefont {Marquardt}}, \bibinfo
		{author} {\bibfnamefont {S.~M.}\ \bibnamefont {Girvin}},\ and\ \bibinfo
		{author} {\bibfnamefont {J.~G.~E.}\ \bibnamefont {Harris}},\ }\bibfield
	{title} {\bibinfo {title} {Strong dispersive coupling of a high-finesse
			cavity to a micromechanical membrane},\ }\href
	{https://doi.org/10.1038/nature06715} {\bibfield  {journal} {\bibinfo
			{journal} {Nature}\ }\textbf {\bibinfo {volume} {452}},\ \bibinfo {pages}
		{72} (\bibinfo {year} {2008})}\BibitemShut {NoStop}%
	\bibitem [{\citenamefont {Jayich}\ \emph {et~al.}(2008)\citenamefont {Jayich},
		\citenamefont {Sankey}, \citenamefont {Zwickl}, \citenamefont {Yang},
		\citenamefont {Thompson}, \citenamefont {Girvin}, \citenamefont {Clerk},
		\citenamefont {Marquardt},\ and\ \citenamefont {Harris}}]{Jayich2008}%
	\BibitemOpen
	\bibfield  {author} {\bibinfo {author} {\bibfnamefont {A.~M.}\ \bibnamefont
			{Jayich}}, \bibinfo {author} {\bibfnamefont {J.~C.}\ \bibnamefont {Sankey}},
		\bibinfo {author} {\bibfnamefont {B.~M.}\ \bibnamefont {Zwickl}}, \bibinfo
		{author} {\bibfnamefont {C.}~\bibnamefont {Yang}}, \bibinfo {author}
		{\bibfnamefont {J.~D.}\ \bibnamefont {Thompson}}, \bibinfo {author}
		{\bibfnamefont {S.~M.}\ \bibnamefont {Girvin}}, \bibinfo {author}
		{\bibfnamefont {A.~A.}\ \bibnamefont {Clerk}}, \bibinfo {author}
		{\bibfnamefont {F.}~\bibnamefont {Marquardt}},\ and\ \bibinfo {author}
		{\bibfnamefont {J.~G.~E.}\ \bibnamefont {Harris}},\ }\bibfield  {title}
	{\bibinfo {title} {Dispersive optomechanics: a membrane inside a cavity},\
	}\href {https://doi.org/10.1088/1367-2630/10/9/095008} {\bibfield  {journal}
		{\bibinfo  {journal} {New J. Phys.}\ }\textbf {\bibinfo {volume} {10}},\
		\bibinfo {pages} {095008} (\bibinfo {year} {2008})}\BibitemShut {NoStop}%
	\bibitem [{\citenamefont {Wilson}\ \emph {et~al.}(2009)\citenamefont {Wilson},
		\citenamefont {Regal}, \citenamefont {Papp},\ and\ \citenamefont
		{Kimble}}]{Wilson2009}%
	\BibitemOpen
	\bibfield  {author} {\bibinfo {author} {\bibfnamefont {D.~J.}\ \bibnamefont
			{Wilson}}, \bibinfo {author} {\bibfnamefont {C.~A.}\ \bibnamefont {Regal}},
		\bibinfo {author} {\bibfnamefont {S.~B.}\ \bibnamefont {Papp}},\ and\
		\bibinfo {author} {\bibfnamefont {H.~J.}\ \bibnamefont {Kimble}},\ }\bibfield
	{title} {\bibinfo {title} {Cavity optomechanics with stoichiometric {SiN}
			films},\ }\href {https://doi.org/10.1103/physrevlett.103.207204} {\bibfield
		{journal} {\bibinfo  {journal} {Phys. Rev. Lett.}\ }\textbf {\bibinfo
			{volume} {103}},\ \bibinfo {pages} {207204} (\bibinfo {year}
		{2009})}\BibitemShut {NoStop}%
	\bibitem [{\citenamefont {Karuza}\ \emph {et~al.}(2013)\citenamefont {Karuza},
		\citenamefont {Biancofiore}, \citenamefont {Bawaj}, \citenamefont
		{Molinelli}, \citenamefont {Galassi}, \citenamefont {Natali}, \citenamefont
		{Tombesi}, \citenamefont {Giuseppe},\ and\ \citenamefont
		{Vitali}}]{Karuza2013}%
	\BibitemOpen
	\bibfield  {author} {\bibinfo {author} {\bibfnamefont {M.}~\bibnamefont
			{Karuza}}, \bibinfo {author} {\bibfnamefont {C.}~\bibnamefont {Biancofiore}},
		\bibinfo {author} {\bibfnamefont {M.}~\bibnamefont {Bawaj}}, \bibinfo
		{author} {\bibfnamefont {C.}~\bibnamefont {Molinelli}}, \bibinfo {author}
		{\bibfnamefont {M.}~\bibnamefont {Galassi}}, \bibinfo {author} {\bibfnamefont
			{R.}~\bibnamefont {Natali}}, \bibinfo {author} {\bibfnamefont
			{P.}~\bibnamefont {Tombesi}}, \bibinfo {author} {\bibfnamefont {G.~D.}\
			\bibnamefont {Giuseppe}},\ and\ \bibinfo {author} {\bibfnamefont
			{D.}~\bibnamefont {Vitali}},\ }\bibfield  {title} {\bibinfo {title}
		{Optomechanically induced transparency in a membrane-in-the-middle setup at
			room temperature},\ }\href {https://doi.org/10.1103/physreva.88.013804}
	{\bibfield  {journal} {\bibinfo  {journal} {Phys. Rev. A}\ }\textbf {\bibinfo
			{volume} {88}},\ \bibinfo {pages} {013804} (\bibinfo {year}
		{2013})}\BibitemShut {NoStop}%
	\bibitem [{\citenamefont {Purdy}\ \emph {et~al.}(2013)\citenamefont {Purdy},
		\citenamefont {Peterson},\ and\ \citenamefont {Regal}}]{Purdy2013}%
	\BibitemOpen
	\bibfield  {author} {\bibinfo {author} {\bibfnamefont {T.~P.}\ \bibnamefont
			{Purdy}}, \bibinfo {author} {\bibfnamefont {R.~W.}\ \bibnamefont
			{Peterson}},\ and\ \bibinfo {author} {\bibfnamefont {C.~A.}\ \bibnamefont
			{Regal}},\ }\bibfield  {title} {\bibinfo {title} {Observation of radiation
			pressure shot noise on a macroscopic object},\ }\href
	{https://doi.org/10.1126/science.1231282} {\bibfield  {journal} {\bibinfo
			{journal} {Science}\ }\textbf {\bibinfo {volume} {339}},\ \bibinfo {pages}
		{801} (\bibinfo {year} {2013})}\BibitemShut {NoStop}%
	\bibitem [{\citenamefont {Reinhardt}\ \emph {et~al.}(2016)\citenamefont
		{Reinhardt}, \citenamefont {M\"uller}, \citenamefont {Bourassa},\ and\
		\citenamefont {Sankey}}]{Reinhardt2016}%
	\BibitemOpen
	\bibfield  {author} {\bibinfo {author} {\bibfnamefont {C.}~\bibnamefont
			{Reinhardt}}, \bibinfo {author} {\bibfnamefont {T.}~\bibnamefont {M\"uller}},
		\bibinfo {author} {\bibfnamefont {A.}~\bibnamefont {Bourassa}},\ and\
		\bibinfo {author} {\bibfnamefont {J.~C.}\ \bibnamefont {Sankey}},\ }\bibfield
	{title} {\bibinfo {title} {Ultralow-noise {SiN} trampoline resonators for
			sensing and optomechanics},\ }\href
	{https://doi.org/10.1103/PhysRevX.6.021001} {\bibfield  {journal} {\bibinfo
			{journal} {Phys. Rev. X}\ }\textbf {\bibinfo {volume} {6}},\ \bibinfo {pages}
		{021001} (\bibinfo {year} {2016})}\BibitemShut {NoStop}%
	\bibitem [{\citenamefont {Hunger}\ \emph {et~al.}(2010)\citenamefont {Hunger},
		\citenamefont {Steinmetz}, \citenamefont {Colombe}, \citenamefont {Deutsch},
		\citenamefont {Hänsch},\ and\ \citenamefont {Reichel}}]{Hunger2010}%
	\BibitemOpen
	\bibfield  {author} {\bibinfo {author} {\bibfnamefont {D.}~\bibnamefont
			{Hunger}}, \bibinfo {author} {\bibfnamefont {T.}~\bibnamefont {Steinmetz}},
		\bibinfo {author} {\bibfnamefont {Y.}~\bibnamefont {Colombe}}, \bibinfo
		{author} {\bibfnamefont {C.}~\bibnamefont {Deutsch}}, \bibinfo {author}
		{\bibfnamefont {T.~W.}\ \bibnamefont {Hänsch}},\ and\ \bibinfo {author}
		{\bibfnamefont {J.}~\bibnamefont {Reichel}},\ }\bibfield  {title} {\bibinfo
		{title} {A fiber fabry{\textendash}perot cavity with high finesse},\ }\href
	{https://doi.org/10.1088/1367-2630/12/6/065038} {\bibfield  {journal}
		{\bibinfo  {journal} {New J. Phys.}\ }\textbf {\bibinfo {volume} {12}},\
		\bibinfo {pages} {065038} (\bibinfo {year} {2010})}\BibitemShut {NoStop}%
	\bibitem [{\citenamefont {Waldron}(1960)}]{Waldron1960}%
	\BibitemOpen
	\bibfield  {author} {\bibinfo {author} {\bibfnamefont {R.}~\bibnamefont
			{Waldron}},\ }\bibfield  {title} {\bibinfo {title} {Perturbation theory of
			resonant cavities},\ }\href {https://doi.org/10.1049/pi-c.1960.0041}
	{\bibfield  {journal} {\bibinfo  {journal} {Proc. {IEE} Part C: Monographs}\
		}\textbf {\bibinfo {volume} {107}},\ \bibinfo {pages} {272} (\bibinfo {year}
		{1960})}\BibitemShut {NoStop}%
	\bibitem [{\citenamefont {Muller}\ \emph {et~al.}(2010)\citenamefont {Muller},
		\citenamefont {Flagg}, \citenamefont {Lawall},\ and\ \citenamefont
		{Solomon}}]{Muller2010}%
	\BibitemOpen
	\bibfield  {author} {\bibinfo {author} {\bibfnamefont {A.}~\bibnamefont
			{Muller}}, \bibinfo {author} {\bibfnamefont {E.~B.}\ \bibnamefont {Flagg}},
		\bibinfo {author} {\bibfnamefont {J.~R.}\ \bibnamefont {Lawall}},\ and\
		\bibinfo {author} {\bibfnamefont {G.~S.}\ \bibnamefont {Solomon}},\
	}\bibfield  {title} {\bibinfo {title} {Ultrahigh-finesse, low-mode-volume
			fabry{\textendash}perot microcavity},\ }\href
	{https://doi.org/10.1364/ol.35.002293} {\bibfield  {journal} {\bibinfo
			{journal} {Opt. Lett.}\ }\textbf {\bibinfo {volume} {35}},\ \bibinfo {pages}
		{2293} (\bibinfo {year} {2010})}\BibitemShut {NoStop}%
	\bibitem [{\citenamefont {Gallego}\ \emph {et~al.}(2016)\citenamefont
		{Gallego}, \citenamefont {Ghosh}, \citenamefont {Alavi}, \citenamefont {Alt},
		\citenamefont {Martinez-Dorantes}, \citenamefont {Meschede},\ and\
		\citenamefont {Ratschbacher}}]{Gallego2016}%
	\BibitemOpen
	\bibfield  {author} {\bibinfo {author} {\bibfnamefont {J.}~\bibnamefont
			{Gallego}}, \bibinfo {author} {\bibfnamefont {S.}~\bibnamefont {Ghosh}},
		\bibinfo {author} {\bibfnamefont {S.~K.}\ \bibnamefont {Alavi}}, \bibinfo
		{author} {\bibfnamefont {W.}~\bibnamefont {Alt}}, \bibinfo {author}
		{\bibfnamefont {M.}~\bibnamefont {Martinez-Dorantes}}, \bibinfo {author}
		{\bibfnamefont {D.}~\bibnamefont {Meschede}},\ and\ \bibinfo {author}
		{\bibfnamefont {L.}~\bibnamefont {Ratschbacher}},\ }\bibfield  {title}
	{\bibinfo {title} {High-finesse fiber fabry{\textendash}perot cavities:
			stabilization and mode matching analysis},\ }\href
	{https://doi.org/10.1007/s00340-015-6281-z} {\bibfield  {journal} {\bibinfo
			{journal} {Appl. Phys. B}\ }\textbf {\bibinfo {volume} {122}},\ \bibinfo
		{pages} {47} (\bibinfo {year} {2016})}\BibitemShut {NoStop}%
	\bibitem [{\citenamefont {Flowers-Jacobs}\ \emph {et~al.}(2012)\citenamefont
		{Flowers-Jacobs}, \citenamefont {Hoch}, \citenamefont {Sankey}, \citenamefont
		{Kashkanova}, \citenamefont {Jayich}, \citenamefont {Deutsch}, \citenamefont
		{Reichel},\ and\ \citenamefont {Harris}}]{Flowers-Jacobs2012}%
	\BibitemOpen
	\bibfield  {author} {\bibinfo {author} {\bibfnamefont {N.~E.}\ \bibnamefont
			{Flowers-Jacobs}}, \bibinfo {author} {\bibfnamefont {S.~W.}\ \bibnamefont
			{Hoch}}, \bibinfo {author} {\bibfnamefont {J.~C.}\ \bibnamefont {Sankey}},
		\bibinfo {author} {\bibfnamefont {A.}~\bibnamefont {Kashkanova}}, \bibinfo
		{author} {\bibfnamefont {A.~M.}\ \bibnamefont {Jayich}}, \bibinfo {author}
		{\bibfnamefont {C.}~\bibnamefont {Deutsch}}, \bibinfo {author} {\bibfnamefont
			{J.}~\bibnamefont {Reichel}},\ and\ \bibinfo {author} {\bibfnamefont
			{J.~G.~E.}\ \bibnamefont {Harris}},\ }\bibfield  {title} {\bibinfo {title}
		{Fiber-cavity-based optomechanical device},\ }\href
	{https://doi.org/10.1063/1.4768779} {\bibfield  {journal} {\bibinfo
			{journal} {Appl. Phys. Lett.}\ }\textbf {\bibinfo {volume} {101}},\ \bibinfo
		{pages} {221109} (\bibinfo {year} {2012})}\BibitemShut {NoStop}%
	\bibitem [{\citenamefont {Brandstätter}\ \emph {et~al.}(2013)\citenamefont
		{Brandstätter}, \citenamefont {{McClung}}, \citenamefont {Schüppert},
		\citenamefont {Casabone}, \citenamefont {Friebe}, \citenamefont {Stute},
		\citenamefont {Schmidt}, \citenamefont {Deutsch}, \citenamefont {Reichel},
		\citenamefont {Blatt},\ and\ \citenamefont {Northup}}]{Brandstatter2013}%
	\BibitemOpen
	\bibfield  {author} {\bibinfo {author} {\bibfnamefont {B.}~\bibnamefont
			{Brandstätter}}, \bibinfo {author} {\bibfnamefont {A.}~\bibnamefont
			{{McClung}}}, \bibinfo {author} {\bibfnamefont {K.}~\bibnamefont
			{Schüppert}}, \bibinfo {author} {\bibfnamefont {B.}~\bibnamefont
			{Casabone}}, \bibinfo {author} {\bibfnamefont {K.}~\bibnamefont {Friebe}},
		\bibinfo {author} {\bibfnamefont {A.}~\bibnamefont {Stute}}, \bibinfo
		{author} {\bibfnamefont {P.~O.}\ \bibnamefont {Schmidt}}, \bibinfo {author}
		{\bibfnamefont {C.}~\bibnamefont {Deutsch}}, \bibinfo {author} {\bibfnamefont
			{J.}~\bibnamefont {Reichel}}, \bibinfo {author} {\bibfnamefont
			{R.}~\bibnamefont {Blatt}},\ and\ \bibinfo {author} {\bibfnamefont {T.~E.}\
			\bibnamefont {Northup}},\ }\bibfield  {title} {\bibinfo {title} {Integrated
			fiber-mirror ion trap for strong ion-cavity coupling},\ }\href
	{https://doi.org/10.1063/1.4838696} {\bibfield  {journal} {\bibinfo
			{journal} {Rev. Sci. Instrum.}\ }\textbf {\bibinfo {volume} {84}},\ \bibinfo
		{pages} {123104} (\bibinfo {year} {2013})}\BibitemShut {NoStop}%
	\bibitem [{\citenamefont {Stapfner}\ \emph {et~al.}(2013)\citenamefont
		{Stapfner}, \citenamefont {Ost}, \citenamefont {Hunger}, \citenamefont
		{Reichel}, \citenamefont {Favero},\ and\ \citenamefont
		{Weig}}]{Stapfner2013}%
	\BibitemOpen
	\bibfield  {author} {\bibinfo {author} {\bibfnamefont {S.}~\bibnamefont
			{Stapfner}}, \bibinfo {author} {\bibfnamefont {L.}~\bibnamefont {Ost}},
		\bibinfo {author} {\bibfnamefont {D.}~\bibnamefont {Hunger}}, \bibinfo
		{author} {\bibfnamefont {J.}~\bibnamefont {Reichel}}, \bibinfo {author}
		{\bibfnamefont {I.}~\bibnamefont {Favero}},\ and\ \bibinfo {author}
		{\bibfnamefont {E.~M.}\ \bibnamefont {Weig}},\ }\bibfield  {title} {\bibinfo
		{title} {Cavity-enhanced optical detection of carbon nanotube brownian
			motion},\ }\href {https://doi.org/10.1063/1.4802746} {\bibfield  {journal}
		{\bibinfo  {journal} {Appl. Phys. Lett.}\ }\textbf {\bibinfo {volume}
			{102}},\ \bibinfo {pages} {151910} (\bibinfo {year} {2013})}\BibitemShut
	{NoStop}%
	\bibitem [{\citenamefont {Kashkanova}\ \emph {et~al.}(2016)\citenamefont
		{Kashkanova}, \citenamefont {Shkarin}, \citenamefont {Brown}, \citenamefont
		{Flowers-Jacobs}, \citenamefont {Childress}, \citenamefont {Hoch},
		\citenamefont {Hohmann}, \citenamefont {Ott}, \citenamefont {Reichel},\ and\
		\citenamefont {Harris}}]{Kashkanova2017}%
	\BibitemOpen
	\bibfield  {author} {\bibinfo {author} {\bibfnamefont {A.~D.}\ \bibnamefont
			{Kashkanova}}, \bibinfo {author} {\bibfnamefont {A.~B.}\ \bibnamefont
			{Shkarin}}, \bibinfo {author} {\bibfnamefont {C.~D.}\ \bibnamefont {Brown}},
		\bibinfo {author} {\bibfnamefont {N.~E.}\ \bibnamefont {Flowers-Jacobs}},
		\bibinfo {author} {\bibfnamefont {L.}~\bibnamefont {Childress}}, \bibinfo
		{author} {\bibfnamefont {S.~W.}\ \bibnamefont {Hoch}}, \bibinfo {author}
		{\bibfnamefont {L.}~\bibnamefont {Hohmann}}, \bibinfo {author} {\bibfnamefont
			{K.}~\bibnamefont {Ott}}, \bibinfo {author} {\bibfnamefont {J.}~\bibnamefont
			{Reichel}},\ and\ \bibinfo {author} {\bibfnamefont {J.~G.~E.}\ \bibnamefont
			{Harris}},\ }\bibfield  {title} {\bibinfo {title} {Superfluid brillouin
			optomechanics},\ }\href {https://doi.org/10.1038/nphys3900} {\bibfield
		{journal} {\bibinfo  {journal} {Nat. Phys.}\ }\textbf {\bibinfo {volume}
			{13}},\ \bibinfo {pages} {74} (\bibinfo {year} {2016})}\BibitemShut {NoStop}%
	\bibitem [{\citenamefont {Fogliano}\ \emph {et~al.}(2021)\citenamefont
		{Fogliano}, \citenamefont {Besga}, \citenamefont {Reigue}, \citenamefont
		{Heringlake}, \citenamefont {Mercier~de Lépinay}, \citenamefont {Vaneph},
		\citenamefont {Reichel}, \citenamefont {Pigeau},\ and\ \citenamefont
		{Arcizet}}]{Fogliano2021}%
	\BibitemOpen
	\bibfield  {author} {\bibinfo {author} {\bibfnamefont {F.}~\bibnamefont
			{Fogliano}}, \bibinfo {author} {\bibfnamefont {B.}~\bibnamefont {Besga}},
		\bibinfo {author} {\bibfnamefont {A.}~\bibnamefont {Reigue}}, \bibinfo
		{author} {\bibfnamefont {P.}~\bibnamefont {Heringlake}}, \bibinfo {author}
		{\bibfnamefont {L.}~\bibnamefont {Mercier~de Lépinay}}, \bibinfo {author}
		{\bibfnamefont {C.}~\bibnamefont {Vaneph}}, \bibinfo {author} {\bibfnamefont
			{J.}~\bibnamefont {Reichel}}, \bibinfo {author} {\bibfnamefont
			{B.}~\bibnamefont {Pigeau}},\ and\ \bibinfo {author} {\bibfnamefont
			{O.}~\bibnamefont {Arcizet}},\ }\bibfield  {title} {\bibinfo {title} {Mapping
			the cavity optomechanical interaction with subwavelength-sized ultrasensitive
			nanomechanical force sensors},\ }\href
	{https://doi.org/10.1103/PhysRevX.11.021009} {\bibfield  {journal} {\bibinfo
			{journal} {Phys. Rev. X}\ }\textbf {\bibinfo {volume} {11}},\ \bibinfo
		{pages} {021009} (\bibinfo {year} {2021})}\BibitemShut {NoStop}%
	\bibitem [{\citenamefont {Norte}\ \emph {et~al.}(2016)\citenamefont {Norte},
		\citenamefont {Moura},\ and\ \citenamefont {Gr\"oblacher}}]{Norte2016}%
	\BibitemOpen
	\bibfield  {author} {\bibinfo {author} {\bibfnamefont {R.~A.}\ \bibnamefont
			{Norte}}, \bibinfo {author} {\bibfnamefont {J.~P.}\ \bibnamefont {Moura}},\
		and\ \bibinfo {author} {\bibfnamefont {S.}~\bibnamefont {Gr\"oblacher}},\
	}\bibfield  {title} {\bibinfo {title} {Mechanical resonators for quantum
			optomechanics experiments at room temperature},\ }\href
	{https://doi.org/10.1103/PhysRevLett.116.147202} {\bibfield  {journal}
		{\bibinfo  {journal} {Phys. Rev. Lett.}\ }\textbf {\bibinfo {volume} {116}},\
		\bibinfo {pages} {147202} (\bibinfo {year} {2016})}\BibitemShut {NoStop}%
	\bibitem [{\citenamefont {Favero}\ and\ \citenamefont
		{Karrai}(2008)}]{Favero2008}%
	\BibitemOpen
	\bibfield  {author} {\bibinfo {author} {\bibfnamefont {I.}~\bibnamefont
			{Favero}}\ and\ \bibinfo {author} {\bibfnamefont {K.}~\bibnamefont
			{Karrai}},\ }\bibfield  {title} {\bibinfo {title} {Cavity cooling of a
			nanomechanical resonator by light scattering},\ }\href
	{https://doi.org/10.1088/1367-2630/10/9/095006} {\bibfield  {journal}
		{\bibinfo  {journal} {New J. Phys.}\ }\textbf {\bibinfo {volume} {10}},\
		\bibinfo {pages} {095006} (\bibinfo {year} {2008})}\BibitemShut {NoStop}%
	\bibitem [{\citenamefont {Tavernarakis}\ \emph {et~al.}(2018)\citenamefont
		{Tavernarakis}, \citenamefont {Stavrinadis}, \citenamefont {Nowak},
		\citenamefont {Tsioutsios}, \citenamefont {Bachtold},\ and\ \citenamefont
		{Verlot}}]{Tavernarakis2018}%
	\BibitemOpen
	\bibfield  {author} {\bibinfo {author} {\bibfnamefont {A.}~\bibnamefont
			{Tavernarakis}}, \bibinfo {author} {\bibfnamefont {A.}~\bibnamefont
			{Stavrinadis}}, \bibinfo {author} {\bibfnamefont {A.}~\bibnamefont {Nowak}},
		\bibinfo {author} {\bibfnamefont {I.}~\bibnamefont {Tsioutsios}}, \bibinfo
		{author} {\bibfnamefont {A.}~\bibnamefont {Bachtold}},\ and\ \bibinfo
		{author} {\bibfnamefont {P.}~\bibnamefont {Verlot}},\ }\bibfield  {title}
	{\bibinfo {title} {Optomechanics with a hybrid carbon nanotube resonator},\
	}\href {https://doi.org/10.1038/s41467-018-03097-z} {\bibfield  {journal}
		{\bibinfo  {journal} {Nat. Commun.}\ }\textbf {\bibinfo {volume} {9}},\
		\bibinfo {pages} {662} (\bibinfo {year} {2018})}\BibitemShut {NoStop}%
	\bibitem [{\citenamefont {Barnard}\ \emph {et~al.}(2019)\citenamefont
		{Barnard}, \citenamefont {Zhang}, \citenamefont {Wiederhecker}, \citenamefont
		{Lipson},\ and\ \citenamefont {McEuen}}]{Barnard2019}%
	\BibitemOpen
	\bibfield  {author} {\bibinfo {author} {\bibfnamefont {A.~W.}\ \bibnamefont
			{Barnard}}, \bibinfo {author} {\bibfnamefont {M.}~\bibnamefont {Zhang}},
		\bibinfo {author} {\bibfnamefont {G.~S.}\ \bibnamefont {Wiederhecker}},
		\bibinfo {author} {\bibfnamefont {M.}~\bibnamefont {Lipson}},\ and\ \bibinfo
		{author} {\bibfnamefont {P.~L.}\ \bibnamefont {McEuen}},\ }\bibfield  {title}
	{\bibinfo {title} {Real-time vibrations of a carbon nanotube},\ }\href
	{https://doi.org/10.1038/s41586-018-0861-0} {\bibfield  {journal} {\bibinfo
			{journal} {Nature}\ }\textbf {\bibinfo {volume} {566}},\ \bibinfo {pages}
		{89} (\bibinfo {year} {2019})}\BibitemShut {NoStop}%
	\bibitem [{\citenamefont {Singh}\ \emph {et~al.}(2014)\citenamefont {Singh},
		\citenamefont {Bosman}, \citenamefont {Schneider}, \citenamefont {Blanter},
		\citenamefont {Castellanos-Gomez},\ and\ \citenamefont {Steele}}]{Singh2014}%
	\BibitemOpen
	\bibfield  {author} {\bibinfo {author} {\bibfnamefont {V.}~\bibnamefont
			{Singh}}, \bibinfo {author} {\bibfnamefont {S.~J.}\ \bibnamefont {Bosman}},
		\bibinfo {author} {\bibfnamefont {B.~H.}\ \bibnamefont {Schneider}}, \bibinfo
		{author} {\bibfnamefont {Y.~M.}\ \bibnamefont {Blanter}}, \bibinfo {author}
		{\bibfnamefont {A.}~\bibnamefont {Castellanos-Gomez}},\ and\ \bibinfo
		{author} {\bibfnamefont {G.~A.}\ \bibnamefont {Steele}},\ }\bibfield  {title}
	{\bibinfo {title} {Optomechanical coupling between a multilayer graphene
			mechanical resonator and a superconducting microwave cavity},\ }\href
	{https://doi.org/10.1038/nnano.2014.168} {\bibfield  {journal} {\bibinfo
			{journal} {Nat. Nanotechnol.}\ }\textbf {\bibinfo {volume} {9}},\ \bibinfo
		{pages} {820} (\bibinfo {year} {2014})}\BibitemShut {NoStop}%
	\bibitem [{\citenamefont {Drever}\ \emph {et~al.}(1983)\citenamefont {Drever},
		\citenamefont {Hall}, \citenamefont {Kowalski}, \citenamefont {Hough},
		\citenamefont {Ford}, \citenamefont {Munley},\ and\ \citenamefont
		{Ward}}]{Drever1983}%
	\BibitemOpen
	\bibfield  {author} {\bibinfo {author} {\bibfnamefont {R.~W.~P.}\
			\bibnamefont {Drever}}, \bibinfo {author} {\bibfnamefont {J.~L.}\
			\bibnamefont {Hall}}, \bibinfo {author} {\bibfnamefont {F.~V.}\ \bibnamefont
			{Kowalski}}, \bibinfo {author} {\bibfnamefont {J.}~\bibnamefont {Hough}},
		\bibinfo {author} {\bibfnamefont {G.~M.}\ \bibnamefont {Ford}}, \bibinfo
		{author} {\bibfnamefont {A.~J.}\ \bibnamefont {Munley}},\ and\ \bibinfo
		{author} {\bibfnamefont {H.}~\bibnamefont {Ward}},\ }\bibfield  {title}
	{\bibinfo {title} {Laser phase and frequency stabilization using an optical
			resonator},\ }\href {https://doi.org/10.1007/bf00702605} {\bibfield
		{journal} {\bibinfo  {journal} {Appl. Phys. B}\ }\textbf {\bibinfo {volume}
			{31}},\ \bibinfo {pages} {97} (\bibinfo {year} {1983})}\BibitemShut {NoStop}%
	\bibitem [{\citenamefont {Ricci}\ \emph {et~al.}(1995)\citenamefont {Ricci},
		\citenamefont {Weidemüller}, \citenamefont {Esslinger}, \citenamefont
		{Hemmerich}, \citenamefont {Zimmermann}, \citenamefont {Vuletic},
		\citenamefont {König},\ and\ \citenamefont {Hänsch}}]{Ricci1995}%
	\BibitemOpen
	\bibfield  {author} {\bibinfo {author} {\bibfnamefont {L.}~\bibnamefont
			{Ricci}}, \bibinfo {author} {\bibfnamefont {M.}~\bibnamefont {Weidemüller}},
		\bibinfo {author} {\bibfnamefont {T.}~\bibnamefont {Esslinger}}, \bibinfo
		{author} {\bibfnamefont {A.}~\bibnamefont {Hemmerich}}, \bibinfo {author}
		{\bibfnamefont {C.}~\bibnamefont {Zimmermann}}, \bibinfo {author}
		{\bibfnamefont {V.}~\bibnamefont {Vuletic}}, \bibinfo {author} {\bibfnamefont
			{W.}~\bibnamefont {König}},\ and\ \bibinfo {author} {\bibfnamefont
			{T.}~\bibnamefont {Hänsch}},\ }\bibfield  {title} {\bibinfo {title} {A
			compact grating-stabilized diode laser system for atomic physics},\ }\href
	{https://doi.org/10.1016/0030-4018(95)00146-y} {\bibfield  {journal}
		{\bibinfo  {journal} {Opt. Commun.}\ }\textbf {\bibinfo {volume} {117}},\
		\bibinfo {pages} {541} (\bibinfo {year} {1995})}\BibitemShut {NoStop}%
	\bibitem [{\citenamefont {Siegman}(1986)}]{Siegman1986}%
	\BibitemOpen
	\bibfield  {author} {\bibinfo {author} {\bibfnamefont {A.~E.}\ \bibnamefont
			{Siegman}},\ }\href@noop {} {\emph {\bibinfo {title} {Lasers}}}\ (\bibinfo
	{publisher} {University Science Books},\ \bibinfo {address} {Mill Valley,
		Calif},\ \bibinfo {year} {1986})\BibitemShut {NoStop}%
	\bibitem [{\citenamefont {Katsidis}\ and\ \citenamefont
		{Siapkas}(2002)}]{Katsidis2002}%
	\BibitemOpen
	\bibfield  {author} {\bibinfo {author} {\bibfnamefont {C.~C.}\ \bibnamefont
			{Katsidis}}\ and\ \bibinfo {author} {\bibfnamefont {D.~I.}\ \bibnamefont
			{Siapkas}},\ }\bibfield  {title} {\bibinfo {title} {General transfer-matrix
			method for optical multilayer systems with coherent, partially coherent, and
			incoherent interference},\ }\href {https://doi.org/10.1364/AO.41.003978}
	{\bibfield  {journal} {\bibinfo  {journal} {Appl. Opt.}\ }\textbf {\bibinfo
			{volume} {41}},\ \bibinfo {pages} {3978} (\bibinfo {year}
		{2002})}\BibitemShut {NoStop}%
	\bibitem [{\citenamefont {Kim}\ \emph {et~al.}(2006)\citenamefont {Kim},
		\citenamefont {Yoon}, \citenamefont {Jang}, \citenamefont {Suh},
		\citenamefont {Kim},\ and\ \citenamefont {Yoon}}]{Kim2006}%
	\BibitemOpen
	\bibfield  {author} {\bibinfo {author} {\bibfnamefont {D.~S.}\ \bibnamefont
			{Kim}}, \bibinfo {author} {\bibfnamefont {S.~G.}\ \bibnamefont {Yoon}},
		\bibinfo {author} {\bibfnamefont {G.~E.}\ \bibnamefont {Jang}}, \bibinfo
		{author} {\bibfnamefont {S.~J.}\ \bibnamefont {Suh}}, \bibinfo {author}
		{\bibfnamefont {H.}~\bibnamefont {Kim}},\ and\ \bibinfo {author}
		{\bibfnamefont {D.~H.}\ \bibnamefont {Yoon}},\ }\bibfield  {title} {\bibinfo
		{title} {Refractive index properties of {SiN} thin films and fabrication of
			{SiN} optical waveguide},\ }\href {https://doi.org/10.1007/s10832-006-9710-x}
	{\bibfield  {journal} {\bibinfo  {journal} {J. Electroceram.}\ }\textbf
		{\bibinfo {volume} {17}},\ \bibinfo {pages} {315} (\bibinfo {year}
		{2006})}\BibitemShut {NoStop}%
	\bibitem [{\citenamefont {Karuza}\ \emph {et~al.}(2012)\citenamefont {Karuza},
		\citenamefont {Molinelli}, \citenamefont {Galassi}, \citenamefont
		{Biancofiore}, \citenamefont {Natali}, \citenamefont {Tombesi}, \citenamefont
		{Giuseppe},\ and\ \citenamefont {Vitali}}]{Karuza2012}%
	\BibitemOpen
	\bibfield  {author} {\bibinfo {author} {\bibfnamefont {M.}~\bibnamefont
			{Karuza}}, \bibinfo {author} {\bibfnamefont {C.}~\bibnamefont {Molinelli}},
		\bibinfo {author} {\bibfnamefont {M.}~\bibnamefont {Galassi}}, \bibinfo
		{author} {\bibfnamefont {C.}~\bibnamefont {Biancofiore}}, \bibinfo {author}
		{\bibfnamefont {R.}~\bibnamefont {Natali}}, \bibinfo {author} {\bibfnamefont
			{P.}~\bibnamefont {Tombesi}}, \bibinfo {author} {\bibfnamefont {G.~D.}\
			\bibnamefont {Giuseppe}},\ and\ \bibinfo {author} {\bibfnamefont
			{D.}~\bibnamefont {Vitali}},\ }\bibfield  {title} {\bibinfo {title}
		{Optomechanical sideband cooling of a thin membrane within a cavity},\ }\href
	{https://doi.org/10.1088/1367-2630/14/9/095015} {\bibfield  {journal}
		{\bibinfo  {journal} {New J. Phys.}\ }\textbf {\bibinfo {volume} {14}},\
		\bibinfo {pages} {095015} (\bibinfo {year} {2012})}\BibitemShut {NoStop}%
	\bibitem [{\citenamefont {Biancofiore}\ \emph {et~al.}(2011)\citenamefont
		{Biancofiore}, \citenamefont {Karuza}, \citenamefont {Galassi}, \citenamefont
		{Natali}, \citenamefont {Tombesi}, \citenamefont {Giuseppe},\ and\
		\citenamefont {Vitali}}]{Biancofiore2011}%
	\BibitemOpen
	\bibfield  {author} {\bibinfo {author} {\bibfnamefont {C.}~\bibnamefont
			{Biancofiore}}, \bibinfo {author} {\bibfnamefont {M.}~\bibnamefont {Karuza}},
		\bibinfo {author} {\bibfnamefont {M.}~\bibnamefont {Galassi}}, \bibinfo
		{author} {\bibfnamefont {R.}~\bibnamefont {Natali}}, \bibinfo {author}
		{\bibfnamefont {P.}~\bibnamefont {Tombesi}}, \bibinfo {author} {\bibfnamefont
			{G.~D.}\ \bibnamefont {Giuseppe}},\ and\ \bibinfo {author} {\bibfnamefont
			{D.}~\bibnamefont {Vitali}},\ }\bibfield  {title} {\bibinfo {title} {Quantum
			dynamics of an optical cavity coupled to a thin semitransparent membrane:
			Effect of membrane absorption},\ }\href
	{https://doi.org/10.1103/physreva.84.033814} {\bibfield  {journal} {\bibinfo
			{journal} {Phys. Rev. A}\ }\textbf {\bibinfo {volume} {84}},\ \bibinfo
		{pages} {033814} (\bibinfo {year} {2011})}\BibitemShut {NoStop}%
	\bibitem [{\citenamefont {Zwickl}\ \emph {et~al.}(2008)\citenamefont {Zwickl},
		\citenamefont {Shanks}, \citenamefont {Jayich}, \citenamefont {Yang},
		\citenamefont {Bleszynski~Jayich}, \citenamefont {Thompson},\ and\
		\citenamefont {Harris}}]{Zwickl2008}%
	\BibitemOpen
	\bibfield  {author} {\bibinfo {author} {\bibfnamefont {B.~M.}\ \bibnamefont
			{Zwickl}}, \bibinfo {author} {\bibfnamefont {W.~E.}\ \bibnamefont {Shanks}},
		\bibinfo {author} {\bibfnamefont {A.~M.}\ \bibnamefont {Jayich}}, \bibinfo
		{author} {\bibfnamefont {C.}~\bibnamefont {Yang}}, \bibinfo {author}
		{\bibfnamefont {A.~C.}\ \bibnamefont {Bleszynski~Jayich}}, \bibinfo {author}
		{\bibfnamefont {J.~D.}\ \bibnamefont {Thompson}},\ and\ \bibinfo {author}
		{\bibfnamefont {J.~G.~E.}\ \bibnamefont {Harris}},\ }\bibfield  {title}
	{\bibinfo {title} {High quality mechanical and optical properties of
			commercial silicon nitride membranes},\ }\href
	{https://doi.org/10.1063/1.2884191} {\bibfield  {journal} {\bibinfo
			{journal} {Appl. Phys. Lett.}\ }\textbf {\bibinfo {volume} {92}},\ \bibinfo
		{pages} {103125} (\bibinfo {year} {2008})}\BibitemShut {NoStop}%
	\bibitem [{\citenamefont {Marquardt}\ \emph {et~al.}(2007)\citenamefont
		{Marquardt}, \citenamefont {Chen}, \citenamefont {Clerk},\ and\ \citenamefont
		{Girvin}}]{Marquardt2007}%
	\BibitemOpen
	\bibfield  {author} {\bibinfo {author} {\bibfnamefont {F.}~\bibnamefont
			{Marquardt}}, \bibinfo {author} {\bibfnamefont {J.~P.}\ \bibnamefont {Chen}},
		\bibinfo {author} {\bibfnamefont {A.~A.}\ \bibnamefont {Clerk}},\ and\
		\bibinfo {author} {\bibfnamefont {S.~M.}\ \bibnamefont {Girvin}},\ }\bibfield
	{title} {\bibinfo {title} {Quantum theory of cavity-assisted sideband
			cooling of mechanical motion},\ }\href
	{https://doi.org/10.1103/PhysRevLett.99.093902} {\bibfield  {journal}
		{\bibinfo  {journal} {Phys. Rev. Lett.}\ }\textbf {\bibinfo {volume} {99}},\
		\bibinfo {pages} {093902} (\bibinfo {year} {2007})}\BibitemShut {NoStop}%
	\bibitem [{\citenamefont {Wilson-Rae}\ \emph {et~al.}(2007)\citenamefont
		{Wilson-Rae}, \citenamefont {Nooshi}, \citenamefont {Zwerger},\ and\
		\citenamefont {Kippenberg}}]{WilsonRae2007}%
	\BibitemOpen
	\bibfield  {author} {\bibinfo {author} {\bibfnamefont {I.}~\bibnamefont
			{Wilson-Rae}}, \bibinfo {author} {\bibfnamefont {N.}~\bibnamefont {Nooshi}},
		\bibinfo {author} {\bibfnamefont {W.}~\bibnamefont {Zwerger}},\ and\ \bibinfo
		{author} {\bibfnamefont {T.~J.}\ \bibnamefont {Kippenberg}},\ }\bibfield
	{title} {\bibinfo {title} {Theory of ground state cooling of a mechanical
			oscillator using dynamical backaction},\ }\href
	{https://doi.org/10.1103/PhysRevLett.99.093901} {\bibfield  {journal}
		{\bibinfo  {journal} {Phys. Rev. Lett.}\ }\textbf {\bibinfo {volume} {99}},\
		\bibinfo {pages} {093901} (\bibinfo {year} {2007})}\BibitemShut {NoStop}%
	\bibitem [{\citenamefont {Aspelmeyer}\ \emph
		{et~al.}(2014{\natexlab{b}})\citenamefont {Aspelmeyer}, \citenamefont
		{Kippenberg},\ and\ \citenamefont {Marquard}}]{Aspelmeyer2014a}%
	\BibitemOpen
	\bibfield  {author} {\bibinfo {author} {\bibfnamefont {M.}~\bibnamefont
			{Aspelmeyer}}, \bibinfo {author} {\bibfnamefont {T.~J.}\ \bibnamefont
			{Kippenberg}},\ and\ \bibinfo {author} {\bibfnamefont {F.}~\bibnamefont
			{Marquard}},\ }\href {https://doi.org/10.1007/978-3-642-55312-7} {\emph
		{\bibinfo {title} {{Cavity Optomechanics - Nano- and Micromechanical
					Resonators Interacting with Light}}}}\ (\bibinfo  {publisher} {Springer
		Berlin Heidelberg},\ \bibinfo {year} {2014})\BibitemShut {NoStop}%
	\bibitem [{\citenamefont {Moser}\ \emph {et~al.}(2013)\citenamefont {Moser},
		\citenamefont {Güttinger}, \citenamefont {Eichler}, \citenamefont
		{Esplandiu}, \citenamefont {Liu}, \citenamefont {Dykman},\ and\ \citenamefont
		{Bachtold}}]{Moser2013}%
	\BibitemOpen
	\bibfield  {author} {\bibinfo {author} {\bibfnamefont {J.}~\bibnamefont
			{Moser}}, \bibinfo {author} {\bibfnamefont {J.}~\bibnamefont {Güttinger}},
		\bibinfo {author} {\bibfnamefont {A.}~\bibnamefont {Eichler}}, \bibinfo
		{author} {\bibfnamefont {M.~J.}\ \bibnamefont {Esplandiu}}, \bibinfo {author}
		{\bibfnamefont {D.~E.}\ \bibnamefont {Liu}}, \bibinfo {author} {\bibfnamefont
			{M.~I.}\ \bibnamefont {Dykman}},\ and\ \bibinfo {author} {\bibfnamefont
			{A.}~\bibnamefont {Bachtold}},\ }\bibfield  {title} {\bibinfo {title}
		{Ultrasensitive force detection with a nanotube mechanical resonator},\
	}\href {https://doi.org/10.1038/nnano.2013.97} {\bibfield  {journal}
		{\bibinfo  {journal} {Nat. Nanotechnol.}\ }\textbf {\bibinfo {volume} {8}},\
		\bibinfo {pages} {493} (\bibinfo {year} {2013})}\BibitemShut {NoStop}%
	\bibitem [{roc(2021)}]{rochau2021}%
	\BibitemOpen
	\bibfield  {title} {\bibinfo {title} {The data within this paper are
			available on zenodo.org, doi:}\ }\href
	{https://doi.org/10.5281/zenodo.5005974} {10.5281/zenodo.5005974} (\bibinfo
	{year} {2021})\BibitemShut {NoStop}%
	\bibitem [{\citenamefont {Dorsel}\ \emph {et~al.}(1983)\citenamefont {Dorsel},
		\citenamefont {McCullen}, \citenamefont {Meystre}, \citenamefont {Vignes},\
		and\ \citenamefont {Walther}}]{Dorsel1983}%
	\BibitemOpen
	\bibfield  {author} {\bibinfo {author} {\bibfnamefont {A.}~\bibnamefont
			{Dorsel}}, \bibinfo {author} {\bibfnamefont {J.~D.}\ \bibnamefont
			{McCullen}}, \bibinfo {author} {\bibfnamefont {P.}~\bibnamefont {Meystre}},
		\bibinfo {author} {\bibfnamefont {E.}~\bibnamefont {Vignes}},\ and\ \bibinfo
		{author} {\bibfnamefont {H.}~\bibnamefont {Walther}},\ }\bibfield  {title}
	{\bibinfo {title} {Optical bistability and mirror confinement induced by
			radiation pressure},\ }\href {https://doi.org/10.1103/physrevlett.51.1550}
	{\bibfield  {journal} {\bibinfo  {journal} {Phys. Rev. Lett.}\ }\textbf
		{\bibinfo {volume} {51}},\ \bibinfo {pages} {1550} (\bibinfo {year}
		{1983})}\BibitemShut {NoStop}%
	\bibitem [{\citenamefont {Olivero}\ and\ \citenamefont
		{Longbothum}(1977)}]{Olivero1977}%
	\BibitemOpen
	\bibfield  {author} {\bibinfo {author} {\bibfnamefont {J.}~\bibnamefont
			{Olivero}}\ and\ \bibinfo {author} {\bibfnamefont {R.}~\bibnamefont
			{Longbothum}},\ }\bibfield  {title} {\bibinfo {title} {Empirical fits to the
			voigt line width: A brief review},\ }\href
	{https://doi.org/https://doi.org/10.1016/0022-4073(77)90161-3} {\bibfield
		{journal} {\bibinfo  {journal} {J. Quant. Spectrosc. Radiat. Transf.}\
		}\textbf {\bibinfo {volume} {17}},\ \bibinfo {pages} {233} (\bibinfo {year}
		{1977})}\BibitemShut {NoStop}%
	\bibitem [{\citenamefont {Chen}\ \emph {et~al.}(2015)\citenamefont {Chen},
		\citenamefont {Meng}, \citenamefont {Wang},\ and\ \citenamefont
		{Chen}}]{Chen2015}%
	\BibitemOpen
	\bibfield  {author} {\bibinfo {author} {\bibfnamefont {M.}~\bibnamefont
			{Chen}}, \bibinfo {author} {\bibfnamefont {Z.}~\bibnamefont {Meng}}, \bibinfo
		{author} {\bibfnamefont {J.}~\bibnamefont {Wang}},\ and\ \bibinfo {author}
		{\bibfnamefont {W.}~\bibnamefont {Chen}},\ }\bibfield  {title} {\bibinfo
		{title} {Ultra-narrow linewidth measurement based on voigt profile fitting},\
	}\href {https://doi.org/10.1364/OE.23.006803} {\bibfield  {journal} {\bibinfo
			{journal} {Opt. Express}\ }\textbf {\bibinfo {volume} {23}},\ \bibinfo
		{pages} {6803} (\bibinfo {year} {2015})}\BibitemShut {NoStop}%
\end{thebibliography}
\end{document}